\pdfoutput=1
\documentclass[jou,11pt]{apa6}
\usepackage[utf8]{inputenc}
\usepackage[T1]{fontenc}
\usepackage{amsmath}
\usepackage{amsfonts}
\usepackage{amssymb}
\usepackage[natbibapa]{apacite}
\usepackage{lmodern}
\usepackage{color}
\usepackage{graphicx}
\usepackage[longnamesfirst]{natbib}
\usepackage{geometry}
\usepackage{lscape}
\author{Heiko H. Sch\"utt$^{1,2}$, Lars O. M. Rothkegel$^2$, Hans A.~Trukenbrod$^2$, Sebastian Reich$^2$, Felix~A.~Wichmann$^1$, Ralf Engbert$^2$}
\title{{\bf \Large Likelihood-based Parameter Estimation and Comparison of Dynamical Cognitive Models}}
\shorttitle{Likelihood-based Evaluation of Dynamical Cognitive Models}
\affiliation{$^1$University of T\"ubingen, $^2$University of Potsdam}

\abstract{
Dynamical models of cognition play an increasingly important role in driving theoretical and experimental research in psychology. Therefore, parameter estimation, model analysis and comparison of dynamical models are of essential importance. Here we propose a maximum-likelihood approach for model analysis in a fully dynamical framework that includes time-ordered experimental data. Our methods can be applied to dynamical models for the prediction of discrete behavior (e.g., movement onsets), in particular, we use a dynamical model of saccade generation in scene viewing as a case study for our approach. For this model, the likelihood function can be computed directly by numerical simulation, which enables more efficient parameter estimation including Bayesian inference to obtain reliable estimates and corresponding credible intervals. Using hierarchical models inference is even possible for individual observers. Furthermore, our likelihood approach can be used to compare different models. In our example, the dynamical framework is shown to outperform non-dynamical statistical models. Additionally, the likelihood based evaluation differentiates model variants, which produced indistinguishable predictions on hitherto used statistics. Our results indicate that the likelihood approach is a promising framework for dynamical cognitive models.}

 
\begin{document}
\maketitle

\textbf{© 2017, American Psychological Association. This paper is not the copy of record and may not exactly replicate the final, authoritative version of the article. Please do not copy or cite without authors permission. The final article will be available, upon publication, via its DOI: 10.1037/rev0000068 (http://dx.doi.org/10.1037/rev0000068)}

\section*{Introduction}
The broad class of dynamical cognitive models \citep{Gelder1998} provides a powerful framework for explaining behavioral data. This modelling approach has been particularly successful in sensorimotor control. For example, an early paradigmatic model was proposed by \citet{Haken1985} who introduced coupled non-linear oscillators as a mathematical model for phase transitions in human finger movements. Another general theory was proposed by \cite{Erlhagen2002} who introduced a flexible framework of movement preparation based on dynamical equations for the temporal evolution of neural fields that specify motor actions in space and time. With their decision field theory, \citet{Busemeyer1993} developed a dynamical framework for decision making in uncertain environments. These representative examples indicate the broad range of dynamical models in cognitive science.  

A strength of the dynamical approach is to generate specific predictions including the dependencies between different data-points over time. This however implies that the statistical treatment of dynamical models requires the comparison of model predictions for time-ordered and interdependent data, which complicates parameter identification and model comparison. As a result, dynamical models are often handled with heuristic and approximate methods. Here we discuss an alternative to these heuristic approaches, namely a statistically well-founded analysis based on the likelihood framework.

An important application of the dynamical framework is the modeling of eye movements. Human observers move their eyes three to four times per second to shift gaze to regions of interest within a given visual scene \citep{Yarbus1967,Henderson2003}. Eye movements are important, since high-acuity vision is limited to the fovea, a small region with a spatial extension of about 2 degrees of visual angle \citep{Helmholtz1924,Nicholls2012}. The analysis of fixated regions permits conclusions on the type of features that attract our gaze. For eye movements in natural scenes, \emph{saliency models} concentrate on predicting the fixation density for large datasets \citep{Itti2001}. The density of fixations provides only information where people look regardless of serial order and durations of fixations. This research strategy turned out to be very successful and a range of saliency models was developed to predict fixation density for a given input image \citep{Kienzle2009,Borji2013,Kummerer2015}.  

Recently, there is an increasing interest in cognitive models that produce sequences of fixations, i.e., a \emph{scanpath}, on a natural scene \citep{Borji2014,Engbert2015,LeMeur2015,Zelinsky2008}. Related models aim at a more complete explanation of the cognitive principles underlying the control of attention and eye movements during exploration of natural scenes. Statistical measures include simple statistics like the distribution of saccade lengths and angles between subsequent saccades \citep{Klein1999,Smith2009}, but also more complex spatial statistics that relate image properties to fixation density \citep{Barthelme2013} or to spatial correlation functions \citep{Engbert2015}.

In the traditional approach for the evaluation of scanpath models, researchers typically simulate scanpaths from their models and compare simulated data to experimentally observed scanpaths using a broad range of statistics \citep{LeMeur2013}. The most common statistics are those associated with the observed experimental data (e.g., distributions of saccade angle and saccade amplitudes). Alternative methods are based on comparisons of scanpaths that include string comparison methods based on the Levenshtein distance \citep[for reading]{Levenshtein1966,Malsburg2011} or vector-based methods \citep{Jarodzka2010}. However, each effect and each discriminating statistic for scanpaths evaluates different aspects of the models. Thus, ranking of model performance depends critically on which effects are investigated and which statistics are applied. None of the statistics used so far quantifies the general agreement between models and experimental data in a dynamical framework.

For saccade generation in dynamical cognitive models, a spatiotemporal map of activations \citep{Erlhagen2002} is built-up according to dynamical evolution equations \citep[e.g.,][]{Jackson1992}. When a saccade target is needed, the activation map is read out to generate a target with a probability that equals the relative activation as determined by the map at the time of saccadic selection. We will study a dynamical model of scanpath generation for eye movements in scene viewing  \citep{Engbert2015}. While we focus on this concrete example to illustrate the procedures of model parameter identification and model comparison, the model only serves as a representative example for the broad class of dynamical cognitive models that are developed for the prediction of sequences of discrete motor actions.

In the current study, we investigate the application of the {\sl likelihood function} as a statistical measure of model performance. The likelihood function of a model $M$ is the probability that a given set of experimental data was generated by the model and a corresponding set of model parameters $\boldsymbol\theta$. Therefore, the likelihood function for a given model depends on the data set and the set of model parameter values that specify the model's behavior. The likelihood is the most widely used measure of model performance in mathematical statistics \citep{Bickel1977,Cox2006}. However, because its numerical computation is believed to be difficult, the likelihood is not yet part of the standard toolbox for dynamical models of cognition. Solving likelihood computation for dynamical models of cognition is potentially very important, since likelihood is the starting point for many additional concepts of statistical inference about model parameters and comparisons between different models, including Bayesian inference \citep{Jaynes2003}.

The likelihood can be computed whenever the model can generate the observed data with a certain probability that is non-zero. This is already guaranteed, if the probability for the next datum can be calculated given the previous data and is greater than zero for any observed datum. This means that the likelihood approach can be applied to an extremely broad class of models.

To investigate how the analysis of dynamical models can benefit from the likelihood approach, we demonstrate numerical computations for the recently published \emph{SceneWalk} model of scanpath generation in natural scene viewing \citep{Engbert2015}. The general motivation for modelling human scanpaths is to derive the rules for the sequential deployment of overt attention (i.e., gaze position) in a natural scene-viewing task. 
The \emph{SceneWalk} model starts from a given spatial distribution of fixation positions (an {\sl empirical saliency map}). Thus, we assume to have perfect knowledge about saliency (up to differences between observers). This is not a strong limitation, since the model could easily be combined with one of the successful saliency models \citep[see][for an overview]{Borji2013}. Thus, our modelling goal is to reproduce the key statistics of human scanpaths (e.g., distribution of saccade lengths and spatial correlations) for a given image, when the time-independent 2D distribution of fixation positions is known to a good approximation.

\section*{Likelihood computation for dynamical models}
\subsection*{Definition of likelihood function}
The fundamental theoretical concept for our approach is the likelihood $L_M(\boldsymbol\theta|\operatorname{data})$ of a model $M$ with parameters $\boldsymbol\theta$ given a specific set of experimental data, which is defined as the conditional probability density $f_M$ for observing the data in the context of model $M$ specified by parameters $\boldsymbol\theta$, i.e.,
\begin{equation}
\label{eq:LikelihoodDef}
L_M(\boldsymbol\theta|\operatorname{data}) = f_M(\operatorname{data}|\boldsymbol\theta) \approx \frac{P_M(\operatorname{data}|\boldsymbol\theta)}{(\Delta A)^N} \;.
\end{equation}
In our case, data are given by a sequence of fixations, for which our models shall predict a density one after another. Each of these densities can be approximated by the probabilities to observe the fixations exactly on a discrete grid, divided by the area each gridpoint represents resulting in a denominator of $(\Delta A)^N$ for $N$ fixations. We will stay with this grid approximation to all likelihoods in this article, as many models are themselves defined on grids, including saliency models and the SceneWalk model that we investigate in the current study. The grid approximation simplifies numerical computations, since this probability is always defined and all integrals reduce to summations over grid points. 

Furthermore we set $\Delta A = 1$, measuring area in grid points, which works, because all models that we aim to compare to each other make predictions on the \emph{same} grid of possible fixation locations. Measuring the area in grid independent units (cm, pixels, degrees of visual angle, etc.) in principle enables comparisons between models, which are defined on different grids.  Using a coarser grid implicitly blurs model predictions for eye movement models and a blurring of the final predictions may change performance considerably \citep{Judd2009}. Thus we think it is preferable to convert all model predictions to the same grid making all necessary conversions explicit.

The likelihood quantifies how well a model describes the data and is the most common criterion for model evaluation in mathematical statistics. Therefore maximizing the likelihood  of a given dataset by optimizing model parameters\footnote{We only consider finite dimensional parameters and models in this paper. We know of no non-parametric models for scanpath generation. A non-parametric model increases the complexity of the analysis considerably. If the reader is interested in this there is a broad literature on non-parametric statistics in both Frequentist \citep{Conover1980} and Bayesian statistics \citep{Gershman2012}} is a straightforward approach to model fitting. Applicability of the likelihood approach depends on both the structure and complexity of a model $M$, i.e., whether the likelihood can be computed exactly (analytically or via numerical simulation of the model) or whether we need to introduce further approximations. If it is not practical to compute the likelihood, likelihood-free strategies for parameter estimation and model comparison have been proposed as an alternative (see Discussion).

\subsection*{The likelihood for dynamical models based on discrete observations}
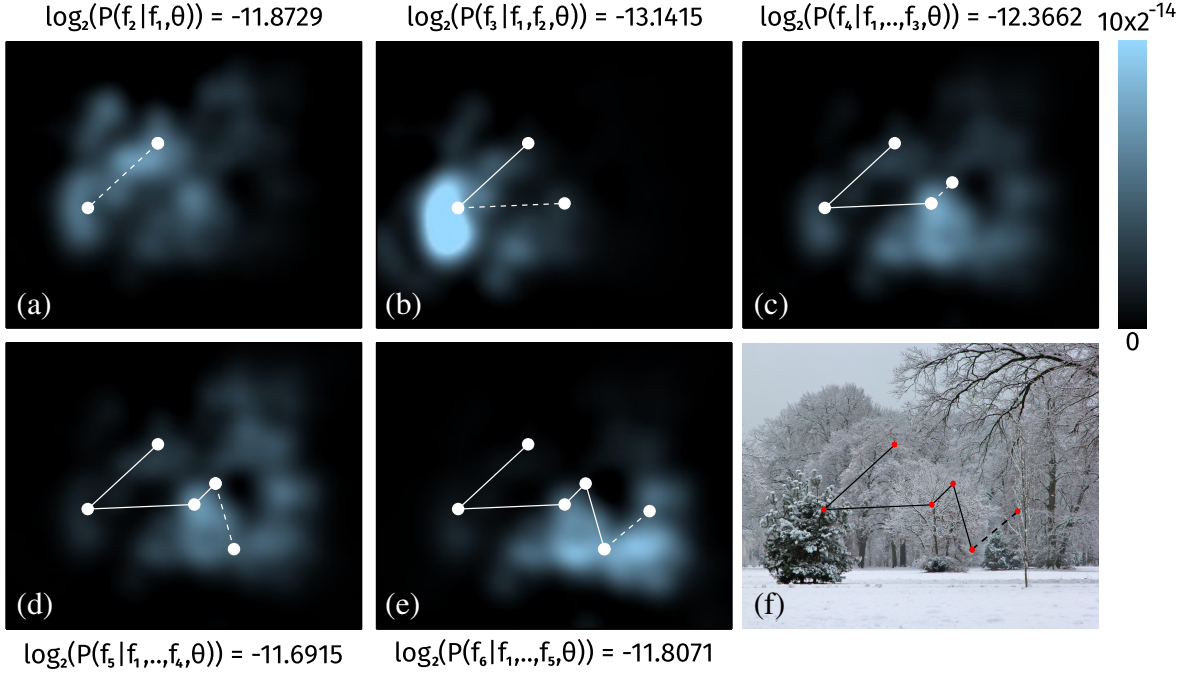
\begin{figure*}[t]
\begin{center}
\unitlength1mm
\begin{picture}(160,90)
\put(0,0){\includegraphics[width=157mm]{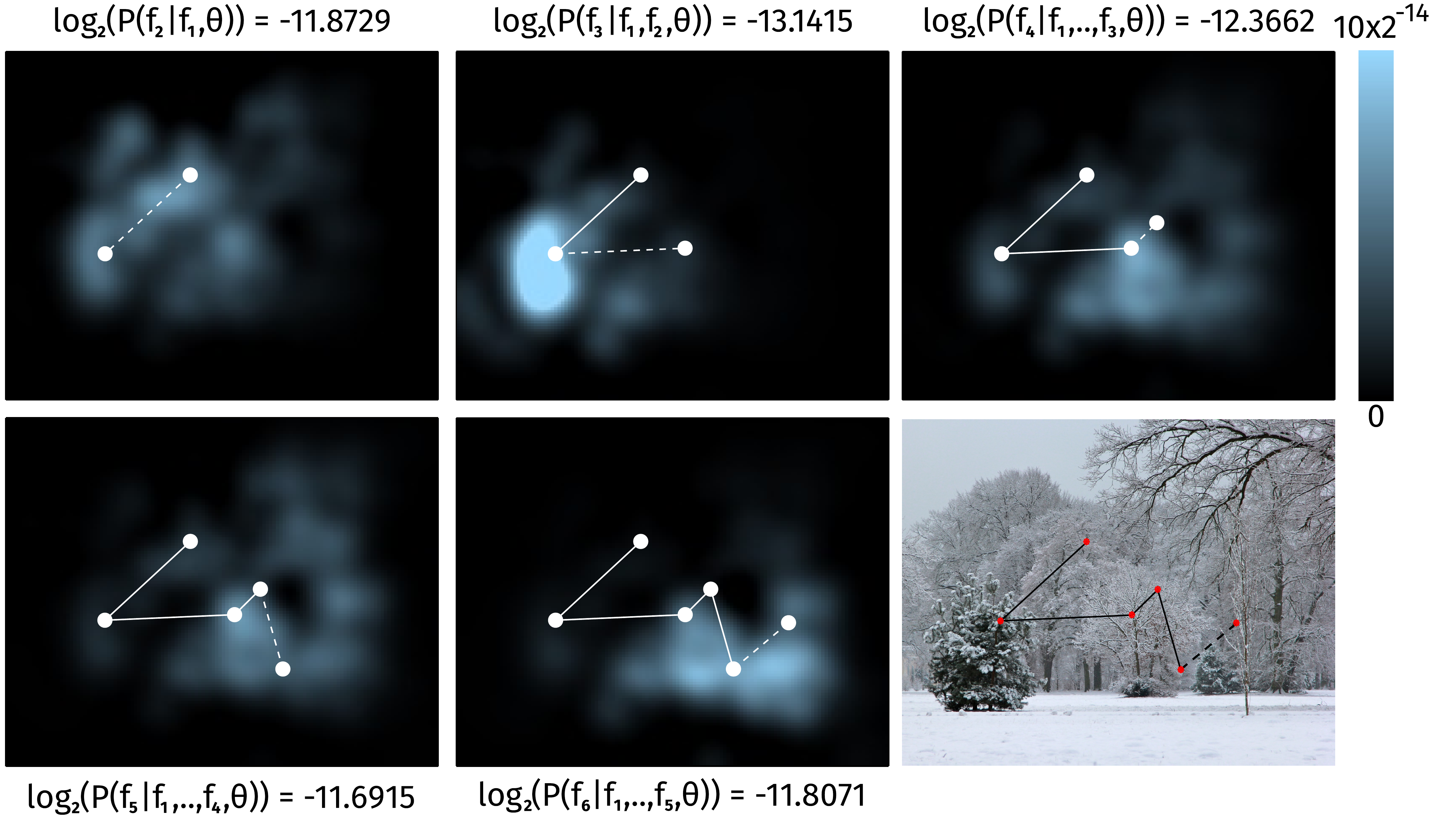}}
\put(2,50){\color{white}\large(a)}
\put(51,50){\color{white}\large(b)}
\put(100,50){\color{white}\large(c)}
\put(2,10){\color{white}\large(d)}
\put(51,10){\color{white}\large(e)}
\put(100,10){\large(f)}
\end{picture}
\end{center}
\caption{\label{fig:LikelihoodModel} Numerical calculation of the likelihood for an example of a fixation sequence. {\bf (a)-(e)} Visualization of the probabilities of the first 5 fixations from a sequence as predicted from the SceneWalk model.
We compute the probability $P(f_i |f_1\dots f_{i-1},\boldsymbol\theta)$ of the next fixation, which the human observer actually generated and force the model to choose the fixation location accordingly. With this new location we can calculate the probability distribution for the next saccade and can thus iterate through the observed scanpaths and calculate their probabilities given by the model and its parameter values. {\bf(f)} The presented image with the scanpath overlayed.} 
\end{figure*}

To calculate the likelihood for dynamical models based on time-ordered experimental data and, specifically, for the SceneWalk model of eye movements in scene viewing \citep{Engbert2015}, we split the likelihood into a product of probabilities for all fixations $f_i=(x_{f_i},y_{f_i})$ given the previous fixations $f_1\dots f_{i-1}$ in the sequence, i.e.,
\begin{equation}
\begin{aligned}
\label{eq:Lcalculation}
L_M(\boldsymbol\theta|\operatorname{data}) &=
L_M(\boldsymbol\theta|f_1,\,f_2,\,\dots,\,f_n)\\ 
&= P_M(f_1) \prod_{i=2}^{n} P_M(f_i|f_1,\,\dots,\,f_{i-1},\boldsymbol\theta) \;,
\end{aligned}
\end{equation}
where $P_M(f_1)$ is the probability of the initial fixation starting at time $t=0$, which can be given by the experimental design or the model. The conditional probabilities $P_M(f_i|f_1\dots f_{i-1},\boldsymbol\theta)$ can be computed by enforcing the model to generate the sequence of fixations $f_1,\,\dots,\,f_{i-1}$ to obtain the probability for the $i^{\rm th}$ fixation $f_i$. This is possible in dynamical models which generate a continuous-time activation map $u$ that translates into a fixation probability $\pi$ to place the next fixation at position $f_i$ at time $t$. Thus, we can read out the probability for the next fixation from the map $u$, Eq.~(\ref{eq:udef}), via the transformation given in Eq.~(\ref{eq:pidef}). During numerical simulation, we force the model to generate a particular scanpath prescribed by the data $f_1,\,f_2,\,\dots$, which translates into a certain probability at each iteration and reduces the necessary computations to a single model run for a given scanpath. This procedure is illustrated for the first fixations on an image in Figure \ref{fig:LikelihoodModel}.

For practical purposes, it is advantageous to use the logarithm of the likelihood (log-likelihood): 
\begin{eqnarray}
l_M(\boldsymbol\theta|{\rm data}))&=&\log(L_M(\boldsymbol\theta|{\rm data}))\\
&=& \sum_{i=1}^N \log(P_M(f_i\boldsymbol|f_1\dots f_{i-1},\boldsymbol\theta))
\end{eqnarray}
The log-likelihood can be calculated and optimized more easily, since it transforms the products over observations into sums of terms and scales numerical values to a more feasible range. 

The log-likelihood characterizes model performance on the whole dataset, in the current case the fixation sequence or scanpath. Therefore, the log-likelihood of a scanpath given a model depends on the length of the sequence or number of fixations. To obtain a number that is easier to compare between different realizations of scanpaths, it is more informative to compute the log-likelihood per fixation, which turns out to represent a sensitive measure of model performance as the log-likelihood is added up over all fixations in a given sequence.

Thus, effectively, we compute the average probability of an observed fixation, calculating the average as a geometric mean. However, we express all likelihoods on a logarithmic scale. When the $\log_2$ is used as we do in this paper, the unit of the log-likelihoods is a \emph{bit}. A difference of 1 bit between two log-likelihood values thus indicates that the corresponding likelihoods differ by a factor of two. 

A log-likelihood of zero indicates that the model predicted the observed data exactly and with probability one. This is a limiting case and certainly not a realistic scenario for typical cognitive models. Almost always models predict a distribution over multiple possible outcomes, which each have smaller probabilities than one. Therefore, log-likelihoods are almost always negative. Indeed the log-likelihoods we calculate below will usually be in the range between $-10\frac{\operatorname{bit}}{\operatorname{fix}}$ and $-20\frac{\operatorname{bit}}{\operatorname{fix}}$.\footnote{Note that these reference values are specific for our choice of grid and area unit, such that they cannot be compared to values obtained with a different grid or area unit. Especially, densities and thus likelihoods can be larger than 1 and log-likelihoods larger than 0, depending on the measure of area chosen.}

\subsection*{Model Details}
\begin{figure*}
\unitlength1mm
\includegraphics[width=\textwidth]{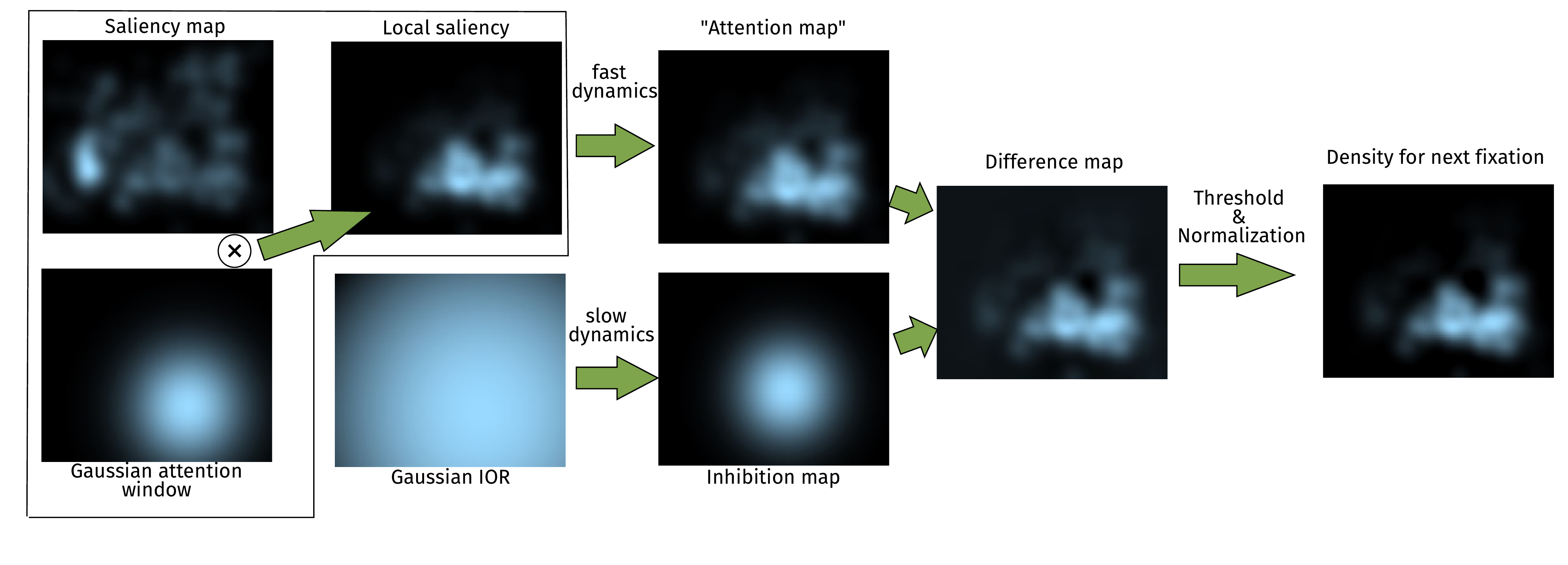}
\caption{\label{fig:ModelIntro}
Schematic illustration of the SceneWalk model \citep{Engbert2015}. The temporal evolution of two independent processing streams for attention and inhibition-of-return is combined into the time-dependent potential $u(x,t)$ that determines the next saccade target. The saliency map is weighted by a Gaussian (attentional window) placed at the current fixation. The resulting local saliency map is used as the input for the build-up of activation in the attention map. An inhibition map is subtracted, which builds up more slowly using a constant-shape Gaussian around the current fixation as input. Finally, thresholding and normalization yield the final distribution $u(x,t)$ for the probabilistic selection of the next saccade target. }
\end{figure*}

For the analysis of the likelihood of the SceneWalk model we need to compute the probability for the next fixation, given all previous fixations in a given trial. In this section we describe how the SceneWalk model computes this probability distributions. To explain this we will require a short recap of the model internals and will take the opportunity to describe the details of some variants of the model we will use to exemplify model comparisons below. 

The SceneWalk model is based on two independent processing streams for excitatory and inhibitory aspects of saccade planning that are related to attentional deployment \citep{Itti1998,Itti2001} and inhibition-of-return \citep{Klein1999,Klein2000}, respectively (Fig.~\ref{fig:ModelIntro}). The excitatory pathway starts with a given fixation density (empirical saliency), which is multiplied with a Gaussian attention window around the current fixation location resulting in a local saliency map. This localization step serves as a first-order approximation to the peripheral loss in available information, cortical processing, and visual attention. For the inhibitory pathway we start with a simple Gaussian around the current fixation marking the currently visited area. The local saliency and the inhibitory Gaussian are both implicitly time-dependent through changes of gaze position.

For a current fixation position $\mathbf{x}_f=(x_f,y_f)$ we first compute the two Gaussian distributions centred at $\mathbf{x}_f$ on a grid of size $L\times L$. The attentional pathway uses a Gaussian aperture $G_A$ with standard deviation $\sigma_A$ to access the static empirical saliency map. The pathway for inhibitory tagging uses a Gaussian $G_F$ with standard deviation $\sigma_F$ to build-up inhibition that drives the model to new regions of the visual field. For a grid position $(x,y)$ these Gaussians are given by
\begin{equation}
\label{eq:Gaussians}
G_{A/F}(x,y; x_f, y_f) = \frac{1}{2\pi\sigma_{A/F}^2}\exp\left(-\frac{(x-x_f)^2+(y-y_f)^2}{2 \sigma_{A/F}^2}\right) \;.
\end{equation}

Next, we define the change over time of the attention map $A(t)=\{A_{ij}(t)\}$ and the fixation map $F(t)=\{F_{ij}(t)\}$ with indices $1\le \{i,j\}\le L$ running over the whole image. Two parameters $\omega_A$ and $\omega_f$ scale the rates of activation change in the two maps and we require the given time-independent salience map $S= \{S_{ij}\}$ and the Gaussians $G_A$ and $G_F$ from equation~(\ref{eq:Gaussians}):
\begin{eqnarray}
\label{Eq:Differential1}
\frac{{\rm d}A_{ij}(t)}{{\rm d}t} =& -\omega_A A_{ij}(t) + \omega_A \frac{S_{ij} \cdot G_A(x_i,y_j;x_f,y_f) }{\sum_{kl} S_{kl} \cdot G_A(x_k,y_l;x_f,y_f)} \\
\label{Eq:Differential2}
 \frac{{\rm d}F_{ij}(t)}{{\rm d}t} =& -\omega_F F_{ij}(t) + \omega_F \frac{G_F(x_i,y_j;x_f,y_f)}{\sum_{kl} G_F(x_k,y_l;x_f,y_f)}  \;,
\end{eqnarray}
where the $\sum_{kl}$ symbol denotes the sum over all grid-points $(k,l)$.

These evolution equations were formulated as difference equations in \cite{Engbert2015}. However, we moved to differential equations here, as they can be solved analytically. By solving  Eqs.~(\ref{Eq:Differential1} \& \ref{Eq:Differential2}), we can exploit the fact that the input $G_{A/F}$ changes only due to saccadic gaze shifts $\mathbf{x}_f\mapsto\mathbf{x^\prime}_f$.  The solution of the differential equations for initial maps $A_0$ and $F_0$ at the start of the fixation at time $t_0$ are given as
\begin{equation}
A(t) = \frac{G_A S}{\sum G_A S} + e^{-\omega_A (t-t_0)} \left(A_0-\frac{G_A S}{\sum G_A S}\right)
\end{equation}
and
\begin{equation}
F(t) = \frac{G_F}{\sum G_F} + e^{-\omega_F (t-t_0)} \left(F_0-\frac{G_F}{\sum G_F}\right) \;,
\end{equation}
where indices have been dropped to simplify the representation. As a consequence of the linear dynamics of the maps, the solutions describe exponential change from the map represented at the beginning of the fixation towards the input map. Using these equations we can calculate the activities at the end of the fixation directly. Another advantage is that this formulation prevents temporal discretization errors \citep[in the original model, a 10~ms temporal discretization was used, see][for details]{Engbert2015}.

At the first fixation the maps in the model need to be initialized. The original model was initialized with zero activities of the maps for attention and inhibitory tagging. For short durations of the first fixation, however, this led to unintended behavior, as the maps are normalized. Small activations on the maps are amplified by the normalization which introduces unwanted starting effects. To prevent this problem of the model's initial conditions, we prepared the maps with a uniform distribution of sum one and adjusted the magnitude of the input such that the equilibrium size of the maps was normalized to one as well. Thus, the sum of activation of the attention map and of the map for inhibitory tagging remains at a constant value of one throughout each simulated trial.

Finally, the two independent activation maps $A(x,t)$ and $F(x,t)$ are combined into a map $u(x,t)$, which is defined as the difference of the attention and inhibition maps after thresholding and normalization. To obtain a flexible relative weighting within each map, numerical values of activations are raised to power $\lambda$ for the attention map $A$ and to power $\gamma$ for the fixation map $F$, respectively. Next, each map is normalized to unit sum \citep{Carandini2012}. Finally, the map for inhibitory tagging is multiplied by a factor $c_{F}$ and subtracted from the attention map. As a result, we obtain a time-dependent potential $u_{ij}(t)$ for target selection:
\begin{equation}
\label{eq:udef}
u_{ij}(t) = \frac{[A_{ij}(t)]^\lambda}{\sum_{kl}[A_{kl}(t)]^\lambda}-c_{F}\frac{[F_{ij}(t)]^\gamma}{\sum_{kl}[F_{kl}(t)]^\gamma}  \;.
\end{equation}
Note that we introduced the factor $c_{F}$ as an additional parameter, which was not present in the original model \citep{Engbert2015}. 

Taking a power of the map at each point changes not only the weighting between different peaks, but also shrinks or widens the individual peaks. Therefore, to obtain parameters which represent the size of the final influence and are thus easier to interpret, we re-parametrized the model using the following equations:

\begin{equation}
\lambda\sigma_A'^2 =\sigma_A^2 \qquad \gamma\sigma_F'^2 =\sigma_F^2
\end{equation}

Thus $\sigma'_A$ and $\sigma'_F$ are the standard deviations the Gaussians would have if they were mapped through the nonlinearity directly.

\emph{Normalization.} To obtain a probability distribution from $u_{ij}(t)$, the potential is normalized to be positive and to have a unit integral over the whole image. Compared to the published version of the model \citep{Engbert2015}, we changed several aspects on the normalization of $u$ and on the initialization of the maps at the beginning of a trial, which are explained in the following. In the normalization procedure of the original model, negative values of the potential $u_{ij}(t)$ implied probability zero to select position $(i,j)$ as the next saccade target. However, this is an unrealistic assumption in the model, since experimental data do not indicate regions which are never selected as a saccade target. We changed the model accordingly. First, we define a function which continuously maps $u$ to an intermediate $u^*$, which is positive everywhere, i.e.,
\begin{equation}
\label{eq:smoothu}
u^*(u) = \left\{ 
\begin{array}{ll}
u  & u>0 \\
0 & u\leq 0
\end{array}
\right.
\end{equation}

In a second step we compute a mixture with a uniform distribution using a weighting factor $\zeta$ to obtain the probability $\pi(i,j)$ for each position on the lattice to be selected as the next fixation target,
\begin{equation}
\label{eq:pidef}
\pi(i,j) = (1-\zeta)\frac{u^*_{ij}}{\sum_{kl} u^*_{kl}} + \zeta \frac{1}{\sum_{kl} 1} \;.
\end{equation}
This formulation maps the original function $u$ to a probability on the map, which always returns a positive probability ($\geq \zeta/\sum_{kl} 1$) for any next fixation. Furthermore, areas with high $u$ are not further distorted by this mapping, such that relative weightings from the original empirical saliency map are kept.

The distribution $\pi(i,j)$ directly represents the probability of a specific grid-point to be the next fixation target, given the previous fixations, i.e., the map to be used in the likelihood calculation described in Equation \ref{eq:Lcalculation} and illustrated in Figure \ref{fig:LikelihoodModel} completing our description of the likelihood calculation for the SceneWalk model.

\subsection*{Competing Models}
Below we will compare the SceneWalk model to some other models, whose details are described in this section.

\emph{Non-dynamic benchmarks}. First, we compare the performance of our model to non-dynamical models that represent limiting cases for saliency evaluation: An image independent spatial bias and empirical saliency. The image independent spatial bias mostly represents the central fixation bias \citep{Buswell1935,Tatler2007}---the experimental observation that observers initially direct their gaze positions toward the image center. A corresponding model can be realized as an image-independent kernel density estimate of all fixations of the full set of images. The empirical saliency model represents the optimal prediction of fixation positions from other observers generated as a kernel density estimate as well, using fixations on the tested image only. Additionally, we implemented a model which generates a uniform distribution over the full image as a null model setting an absolute zero point on our log-likelihood scale.

\emph{A model without inhibition.} As a first dynamical model to compare to, we chose a model without inhibition, to test whether this part of the model is necessary as the influence of inhibition of return on scene viewing behavior has been challenged recently \citep{Smith2009}.
To implement this model we simply set $c_F=0$ in our original model removing the influence of the inhibitory pathway. As $u$ then cannot become negative anymore, we also replaced the mapping from $u$ to $u^*$ with the identity. As a consequence, all parameters of the inhibitory pathway are superfluous in this model, such that we are left with only 4 parameters for this model: $\omega_A,\sigma_A,\lambda$ and $\zeta$.

\emph{Divisive inhibition model.}
The original SceneWalk model implements a subtractive inhibition. However, there are no strong reasons, why this inhibition should be subtractive. An alternative and common model of interaction is divisive inhibition \citep{Carandini2012}. To test this alternative form of combining the two maps, we changed the formula for $u$ to:

\begin{equation}
u_{ij}(t) = \frac{[A_{ij}(t)]^\lambda}{c_{F}^\gamma+[F_{ij}(t)]^\gamma}
\end{equation}

As for the model without inhibition, the variable $u$ cannot become negative. Again, we replaced the mapping from $u$ to $u^*$ with the identity. This way to combine excitation and inhibition has the same number of parameters as the original subtractive formulas. Thus we are left with 8 parameters as for the original model.

\section*{Estimation of model parameters}
As it is common practice our previous approach to the estimation of model parameters was based on minimization of an ad hoc loss function that included gaze positions and saccade lengths as measures of model performance (see Appendix in Engbert et al., 2015). First, we computed the squared differences between densities of gaze positions from experimental and simulated data using 2D bins for discretization. Second, we compared experimentally observed and simulated saccade lengths via squared differences from bins of the distributions. The sum of both measures was minimized to obtain parameter estimates. 

However, there were several problems associated with this approach that motivated us to develop an alternative framework. First, our earlier approach worked for a limited set of parameters only. Some of the parameters had to be fixed at plausible values. These fixed parameters included important parameters, for example, normalization exponents of the dynamic activation maps, which are critical for the spatial correlation functions we intended to reproduce. Second, the qualitative model analyses necessary to find useful and plausible values for the fixed parameters required time-consuming hand-selected model runs. Third, our earlier fitting approach based on a subset of hand-selected fixed parameters and estimates from minimization of an ad-hoc loss-function could not guarantee reliable or consistent estimates and was missing a statistical justification. Moreover, confidence intervals of the model parameters were inaccessible and were, therefore, replaced by an ad-hoc indicator of errors of parameter estimates derived from multiple runs of the minimization algorithm. Due to these shortcomings of the earlier approach, we set out to develop an improved strategy for parameter estimation that would be statistically well-founded, reliable, and efficient in terms of computer time, while working for all parameters.

\subsection*{Maximum likelihood estimation}
A tutorial on the MLE concept for model fitting is given by \cite{Myung2003} in the context of mathematical models in psychology \citep[see][for a more general context]{Hays1994}. The general idea is to find the particular (vector-valued) parameter $\boldsymbol\theta$ that corresponds to the maximum of the likelihood function given the observed data. This parameter value is used as a parameter estimate and, therefore, termed \emph{maximum likelihood estimate} (MLE).

Fitting models to data based on the likelihood has considerable statistical advantages over using other statistics for fitting \citep{Myung2003}. First, the likelihood guarantees sufficiency, i.e., raw data do not constrain the parameters more than the maximum likelihood criterion. Second, for the likelihood, there is asymptotic consistency, such that for large samples the estimate converges to the correct parameter value if the data were generated from the model. Third, the likelihood has asymptotic maximum efficiency, i.e., for large samples, there is no consistent estimate with smaller variance. Finally, the likelihood estimate is not changed by the re-parametrization of the model, which is known as parametrization invariance. 

In numerical simulation models like the SceneWalk model, the maximum of the likelihood can be found using an optimization algorithm that evaluates the likelihood $L_M(\boldsymbol\theta|{\rm data})$ varying the model parameters $\boldsymbol\theta$. Most optimization algorithms try to change the parameters gradually to improve the likelihood and can thus be trapped in local extrema, where the likelihood is higher than for surrounding parameter values, but not the globally best parameter value. If the global optimum is found, it must not depend on the specific optimization algorithm or starting position. Consequently it is common practice to run multiple optimizations with different starting positions. If one of the local extrema is clearly better than the others and the optimizations end up in clusters, one can be reasonably sure that one found the global optimum. 

Alternatively the field of global optimization designs algorithms to find global minima. Two well known families of algorithms for global optimization are: Simulated annealing, which---inspired by the cooling of physical materials---first explores broadly and later allows less and less bad objective values settling near the optimum \citep{Kirkpatrick1983,Kirkpatrick1984}, and the Genetic algorithm, which simulates a population of parameter values over generations in which points with high objective function values have higher probability to reproduce in the next generation \citep{Holland1975,Golberg1989,Houck1995}. Variants of both these algorithms are available for most higher programming languages like MATLAB or python. As a promising idea for the  future the relatively recent meta-modelling approach aims to model our knowledge about the function gained so far and to conclude which points to sample to gain the most information about the optimum \citep{Jones1998,Villemonteix2009,Hennig2012}.

For optimization of the parameters of the SceneWalk model we employed the genetic algorithm for global optimization as implemented in MATLAB (R2016a). We used 200 individuals on the logarithm of the parameters with a range from $-10$ to $10$ corresponding to a range from $0.000\,045$ to $22\,026$ for the parameters. Subsequently we further optimized using the Nelder-Mead Simplex Algorithm as implemented as fminsearch in MATLAB. Using the standard settings for all other options these algorithms found the global maximum reliably, as confirmed by some standard optimization runs from random start positions, the sampling we did for Bayesian inference and the fits we computed for cross validation as described below.

\subsection*{Bayesian inference}
If the likelihood $L_M(\boldsymbol\theta|{\rm data})$ of the data can be computed for a given model $M$, then  Bayesian inference \citep[for overviews]{Marin2007,Gelman2014} is a viable method for parameter estimation. The main advantage of Bayesian inference in the current context is that it provides not only the best fitting parameter values, but also a full distribution of possible parameter values. Thus, there is information on which other parameter values could also explain the data and thus how well the parameters of the assumed model are constrained by given data. In Bayesian inference, the goal is the computation of a posterior distribution $P(\boldsymbol\theta|{\rm data})$ that indicates the most probable parameter values $\boldsymbol\theta$ under the assumption of model $M$ and the given data. Based on the likelihood $L_M(\boldsymbol\theta|{\rm data})$ and a prior distribution $P(\boldsymbol\theta)$, which describes our knowledge or beliefs about the parameters prior to data collection, the posterior distribution is computed as
\begin{equation}
\label{eq:Bayes}
P(\boldsymbol\theta|{\rm data}) = \frac{L(\boldsymbol\theta|{\rm data})P(\boldsymbol\theta)}{\int_\Omega P(\boldsymbol\theta)L(\boldsymbol\theta|{\rm data})d\boldsymbol\theta} \;,
\end{equation}
where, computationally, the main problem is that quantities of interest are usually integrals over the posterior $P(\boldsymbol\theta|{\rm data})$ like the expected value of the posterior, its variances or correlations. To compute these integrals it is often necessary to use Markov Chain Monte Carlo (MCMC) methods \citep{Brooks2011,Robert2013}. These methods produce---sometimes weighted---samples from the posterior using only local evaluations of the likelihood and prior. These samples can then be used to replace integrals by sample means. This especially avoids the direct calculation of the denominator $P(\operatorname{data})= \int_\Omega P(\boldsymbol\theta)L(\boldsymbol\theta|{\rm data})d\boldsymbol\theta$, which in turn can be computed from the samples if one is interested in this value.  

The most controversial aspect of Bayesian statistics is the choice of prior. The main reason is that the prior may serve very different functions in different situations.

The first most literal interpretation of priors is that they shall represent all available believes prior to the experiment. If one manages to formulate all prior believes into the prior distribution, the posterior represents the believes one should have after the experiment to do proper reasoning \citep[Chapter 1]{Jaynes2003}. If we had an estimate of the parameters from some other experiment, or had any other kind of information what the parameters or predictions of the model should be, the prior offers a possibility to include this knowledge. In the absence of prior information the general recommendation is to use relatively broad uninformative priors to avoid biasing the conclusions  too much. If a bias is unavoidable, then the recommendation is modified to use a prior which favors the opposite of the suspected conclusion to achieve a conservative analysis showing how well the data should convince a sceptic (\citealp[Chapter 2.8]{Gelman2014}, \citealp[Chapter 11 \& 12]{Jaynes2003}).

The notion of an uninformative prior can be formalized mathematically, which leads to Jeffreys' priors \citep{Jeffreys1946}. Another mathematically preferable kind of prior are conjugate priors, for which the posterior has the same form as the prior \citep[Chapter 2.4]{Gelman2014} such that posteriors can be parametrized and analytically analyzed. Neither Jeffreys' priors nor conjugate priors are particularly relevant for the complex models we study here, as they are rarely known or even computable for highly complex models.

A second more objective interpretation is that the priors shall represent the actual distribution of parameters as close as possible. In this interpretation, which is popular in machine learning, the prior becomes part of the model to be evaluated. The better the prior represents the distribution of parameters needed to fit data, the better it is. Obviously such evaluations require multiple instances for which a parameter is fitted. Once one starts to adjust the prior to fit some data this approach becomes essentially equivalent to hierarchical models which we discuss below.

Prior assumptions on parameters also represent a helpful tool to include information obtained from other experiments and other knowledge (e.g., physiological constraints) or to \emph{regularize} the model, which is a general expression for preferring some parameter values of the model over others, if both parameter values explain the data equally well. The term regularization is used usually in Frequentist contexts and justified as a means to stabilize model fitting when the parameters are not sufficiently constrained by the data.

For regularization purposes one typically differentiates whether parameter values are only considered less likely or impossible. Only the former is usually called regularization, the later is usually called constrained estimation. This distinction is mainly necessary because once there are areas of parameter space which are impossible the algorithms for optimization or sampling need to be changed. For the effect of the priors on the model this is a more gradual distinction. While it is usually discouraged to entirely exclude parameter values a priori, i.e., to set their prior probability to $0$, very small prior probabilities will have the same effect on the model predictions and parameter fits.

The different aims for priors partially work against each other. To regularize or to include prior knowledge helps mostly if the parameters could not be constrained well by the data at hand, i.e. when the prior excludes parameters which could fit the data convincingly as well. When doing this one can obviously not interpret the posterior as information how well these parameters are constrained by the data. Thus different aims might require different priors for the same model and data.

As we do not require regularization and have little to no prior information about the parameters of the model we investigate, we chose an extremely broad prior not to influence our parameter estimates. We assume a log-normal distribution with a standard deviation of 30 units (log-space) around 0 (in log-space).

\subsection*{Results on model parameter estimation}
For the SceneWalk model, we used the same dataset as in the original article \citep{Engbert2015}. In the experimental data, gaze positions were recorded via eye tracking from 35 human observers in a memorization task. Experimental stimuli consisted of 15 natural images and 15 texture images, where the latter are photographs of relatively homogeneous textures like grass or a stone wall.

The numerical optimization of the model parameters required less computation time than the original fitting method, as the likelihood objective is not stochastic, although we fitted four more parameters (the pooling exponents $\lambda$ and $\gamma$, the weighting of the inhibitory map $c_{F}$ and the weight of the uniform map in the mixture $\zeta$).

The results of the Maximum Likelihood estimation are listed in Table \ref{tab:ParValues}. As they agree with values from Bayesian estimation we shall discuss their meaning after explaining the origin of the Bayesian estimates.

To perform Bayesian inference about the parameters of the SceneWalk model, we sampled the posterior distribution with a  Metropolis Hastings algorithm \citep{Metropolis1953,Hastings1970}. A hand-tuned multivariate Gaussian proposal distribution was chosen to have a covariance matrix roughly proportional to the covariance of the sampled distribution and to reach an acceptance rate of roughly $25\%$ as recommended as optimal for Gaussians by \cite{Gelman1996}. We restricted us to reproduce the diagonal of the covariance Matrix, i.e., to the variances of the individual parameters, and 3 particularly strong covariances, between $\sigma_A$ and $\sigma_F$, $C_F$ and $\lambda$ and $C_F$ and $\zeta$ respectively. Using this scheme we sampled three chains with $50\,000$ samples each starting with a small displacement from the MAP estimate. We then discarded the first 1000 samples as burn in, which covered the initial transient back towards the MAP in all parameters.

First we checked that our sampling algorithm converged using the $\hat{R}$ statistic \citep{Gelman1992,Brooks1998}, which quantifies how large the variance between chains is compared to the variance within the chains, i.e., whether the chains sampled different regions. The $\hat{R}$ statistic is always greater than one and, when the chains under analysis converged to the same stationary distribution, the $\hat{R}$ statistic should be close to one. For our chains we obtained values in the range from $1.00$ to $1.06$ for different parameters and a value of $1.06$, when $\hat{R}$ was computed as a multivariate statistic. We thus concluded that our chains converged to their common stationary distribution, which we also confirmed by investigating visually and by comparison of the distributions obtained from the three independent chains. 

Next we checked that our chains mixed sufficiently well, i.e., we tested that the samples were sufficiently uncorrelated with each other and, therefore, that the samples provide an adequate representation of the posterior distribution. The mixing property was analysed via the effective sample size, which is an estimate of the number of independent samples one would need to get an equally good representation of the posterior. This estimate is computed from the autocorrelation of the chain for each individual parameter. As a result, we obtained an estimate of the effective sample size for each parameter, although the true efficiency of the sampling algorithm is a single quality of the method. For our chains, the effective sample sizes turned out to range from 624 to 22806 for the different parameters. This indicates that our sampling algorithm provides at least the information of a few hundred samples, which we considered as sufficient for our purposes. 

However, our findings on the effective sample size also indicate that the Metropolis Hastings algorithm could probably be improved in efficiency as its sampling efficiency (effective sample size divided by the number of drawn samples) was less than $1\%$. When the algorithm is well tuned to the problem, a sampling efficiency of several percent  can be reached \citep{Gelman1996}.

\begin{figure*}
\vspace{1cm}
\makebox[\textwidth][c]{
\includegraphics[width = .9\textwidth, trim = 1cm 1.5cm 1cm 1cm]{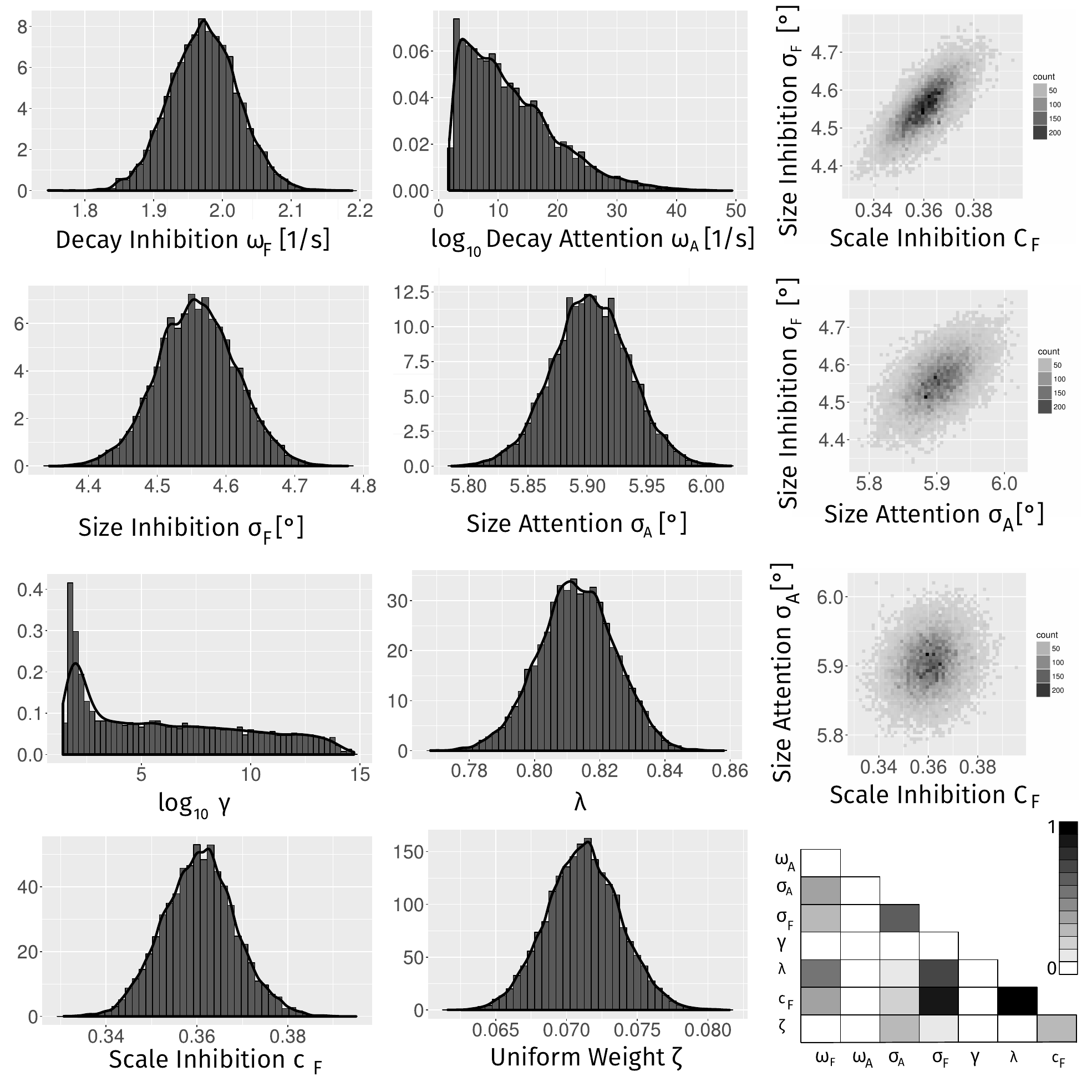}
}
\renewcommand{\baselinestretch}{1.1}
\vspace{1cm}
\caption{Sampling results for the posterior distribution for the example model's parameters. In the left two columns we show histograms and density estimates for all 8 parameters. Except for $\gamma$ and $\omega_A$ all parameters seem to be well constrained by the data. In the right column we show two dimensional histograms of two parameters against each other illustrating their dependencies. The first indicates the strong correlation between the spatial scale and scaling factor of the inhibition. The second shows the medium strength dependency between the sizes of inhibition and attention pathway. The third plot illustrates the near independence of the spatial scale of the attention map and the scaling factor highlighting the non transitivity of correlations. In the lower right corner we present a summary plot about the correlations between parameters. The darkness of each rectangle in this plot indicates the absolute correlation between two parameters, which each could be shown as a 2D histogram as we did for 3 examples above.
\label{fig:sampling}}
\end{figure*}

The sampled posterior distributions are displayed in Figure \ref{fig:sampling}. The distributions clearly indicate the most likely values of the parameters. All parameters except for the decay of the excitatory map $\omega_A$ and the exponent $\gamma$ were well constrained by the data. Their posterior marginals concentrate on a range of $\leq \pm 10\%$ around the best fitting values and are much narrower than the prior ($\pm 10$ log-units).

\begin{table*}
\caption{\normalfont{Table of the parameter values obtained from different point estimates. Displayed are the maximum likelihood estimate (MLE), the posterior mean estimate ($\pm$ its estimated sampling error) and a credible interval from the Bayesian estimation we present, compared to the values from the original study by \citet{Engbert2015}. Values marked with * were fixed without fitting in the original article. \label{tab:ParValues}}}
\vspace{1ex}
\makebox[\textwidth][c]{
\begin{tabular}{c|c|c|rl|c c}
parameter name & original estimate & MLE & \multicolumn{2}{c|}{posterior mean estimate} & \multicolumn{2}{c}{95\% credible interval}\\
\hline
$\omega_A$ &  6.607  &$2.4\times10^{30}$&$1.1\times10^{45}$ &$\pm 8\times10^{44}$ & 417.6 & $4.373\times10^{30}$\\
$\omega_F$ &  0.00903&$1.9298$&$1.973$ &$\pm 0.001601$ & 1.876 & 2.071\\
$\sigma_A$ &   4.88  &$5.9082$&$5.903$ &$\pm 0.000640$ & 5.838 & 5.967\\
$\sigma_F$ &   3.9436&$4.5531$&$4.558$ &$\pm 0.002282$ & 4.445 & 4.671\\
$\gamma$ &     0.3*  &$44.780$&$3.3\times10^{12}$ &$\pm 4.5\times10^{11}$ & 43.83 & $3.249\times 10^{13}$\\
$\lambda$ &    1*    &$0.8115$&$0.8130$&$\pm 0.000422$ & 0.7896 & 0.8354\\
$c_{F}$ & 1*         &$0.3637$&$0.3605$&$\pm 0.000321$ & 0.3658 & 0.3767\\
$\zeta$ &   ---      &$0.0722$&$0.0712$&$\pm 0.000046$ & 0.0662 & 0.0764
\end{tabular}}
\end{table*}

From an analysis of the marginal posterior distributions displayed in Figure~\ref{fig:sampling}, we can extract point estimates and credible intervals, which characterize a single optimal model parameter and a range that contains the true parameter value with a given probability. For our model we extracted the mean estimate and a 95\% credible interval for each parameter listed in Table~\ref{tab:ParValues} to compare them to the parameter estimates obtained in the original paper \citep{Engbert2015}. For the well constrained parameters the MLE and mean estimates agree closely as expected. These estimates can only differ when the posterior is relatively broad. Consequently, our interpretation is the same for both parameter estimates.

Qualitatively, we reproduce the patterns observed in the original paper: The activation on the excitatory attention map is larger and faster than the inhibitory fixation map ($\omega_A>\omega_F$, $\sigma_A>\sigma_F$). Quantitatively, the parameters differ substantially from the ones in the original study. In particular, compared to the original study, (i) the Gaussian input around the current fixation is larger by roughly a degree for both maps, (ii) the inhibitory fixation map is 2.5 log-units faster, the attention map could be arbitrarily fast and (iii) the pooling exponents ($\gamma$\ and $\lambda$) converged to very different values than those chosen by hand. 

The fact that the two parameters $\gamma$ and $\omega_A$  are not well constrained can be explained as follows. The parameter $\omega_A$ determines the rise-rate of the attention map. Once this rate is fast enough, changes of the parameter value will not influence predictions any more. Similarly high values of gamma produce all very similar nonlinearities in the inhibition map and thus do not change any predictions. As we discussed above one could have used a prior to restrict these parameters to ranges over which they change predictions to avoid the result of parameters which are unconstrained over such wide ranges. This would however hide the fact that they are not well constrained from the posterior sampling result.

From the posterior distribution, we can also extract two-dimensional marginal distributions as histograms or density estimates. These marginal distributions illustrate posterior couplings between pairs of parameters. Such couplings indicate that obtaining information of one of the two parameters would constrain both of them better. 
For example, we show two-dimensional histograms for 3 pairs of parameters (Fig.~\ref{fig:sampling}):
\begin{itemize}
\item For $\sigma_F$ and $C_f$ we find a relatively strong coupling which indicates that models with stronger inhibition require it to be spread wider to explain the data equally well. 
\item For $\sigma_A$ and $\sigma_F$ we find a weaker, but still visible coupling, which indicates that the inhibition and attention window need to covary in size to explain the data.
\item Finally, $\sigma_A$ and $C_F$ turned out to be approximately independent. Fixing one of these parameters would not constrain the other parameter.
\end{itemize}
This last point additionally illustrates that posterior correlations are not necessarily transitive.

In summary, the posterior marginal distributions can be reduced to the correlation coefficient, which captures the strength of the linear dependence between the parameters. These correlation coefficients are also plotted in Figure \ref{fig:sampling} for each combination of two parameters. The samples from the posterior also contain all higher-order dependencies between parameters, although they are more difficult to visualize or summarize. 

\subsection{Inter-Subject differences and Hierarchical Models}

\begin{figure*}
\includegraphics[width=\textwidth, trim = 4cm 0 4cm 0]{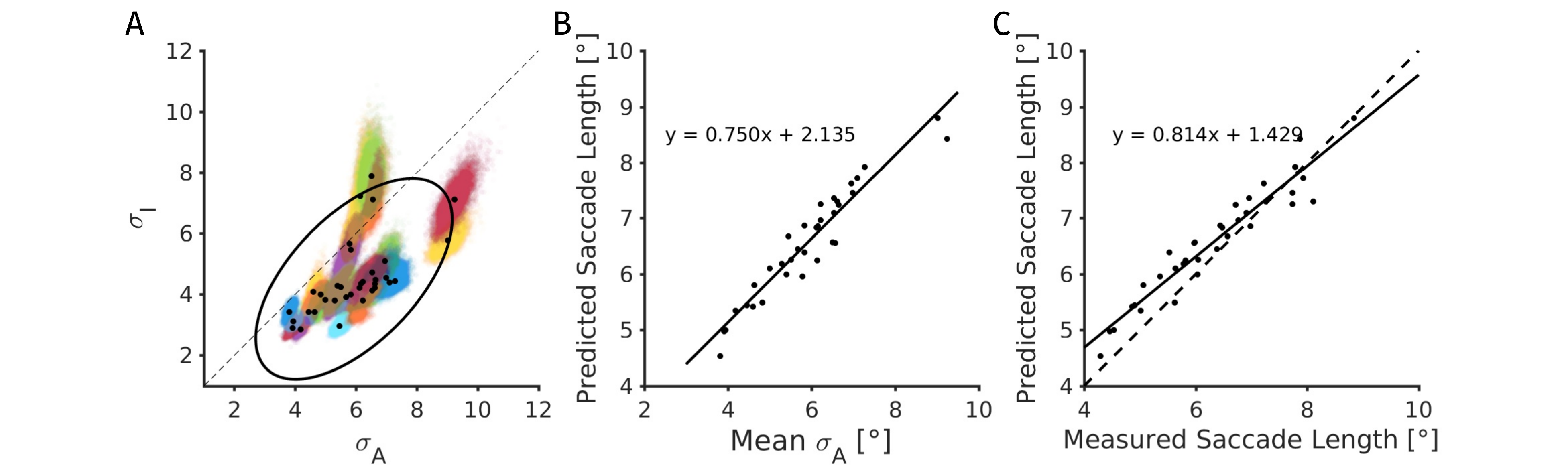}
\caption{Results for the Hierarchical model. \textbf{A}: Fits for the two parameters $\sigma_A$ and $\sigma_F$ for the different observers. Each observer is represented by a black marker marking their posterior mean and a colored point cloud representing the posterior samples. Additionally the dashed line marks the $\sigma_A=\sigma_F$ diagonal and a large black ellipse marks the posterior mean estimate for the 95\% line of the Gaussian population model. \textbf{B}: Predicted saccade length for each subject against their posterior mean estimate for $\sigma_A$ with a linear least squares regression line. \textbf{C}: Predicted mean saccade length from the posterior mean estimate against the measured mean saccade length for each subject. The dashed and continous line mark the equallity diagonal and a linear least squares regression line. \label{fig:Hierarchical}}
\end{figure*}

For many cognitive tasks subjects differ in meaningful ways, which we might want to include into our models. For eye movements, one important subject-specific parameter is the average length of saccades \citep{Castelhano2008}. For our participants who generated the longest saccades, we observed average saccade lengths twice as large as the saccade lengths for participants with the shortest saccades (see Figure \ref{fig:Hierarchical}).

One popular method for integrating differences between subjects into models are hierarchical models. In hierarchical models the differences between subjects are explained by assuming different parameter values for each subject which follow an additional model for the distribution of parameters in the population.\footnote{The hierarchical model framework can also be used to model effects of other properties of the task like item and image effects.} The main advantage of using a model for the distribution of parameters in the population is to stabilize the estimates for subjects, whose parameters are not well constrained by the data alone.

We implemented a hierarchical model which allows the sizes of the attention span and of the inhibited area to differ between subjects in order to explain the observed differences in saccade length. To simplify the analysis we fixed all other parameters of the model to their MAP estimates over all subjects and images from the model fitting explained above. 

As our model for the parameter distribution in the population, we introduced a two dimensional Gaussian, which we parametrized using means and variances for the two parameters and the correlation between parameters as a fifth parameter. As we now aim to estimate these five parameters together with the individual subjects parameters, we defined a prior on each parameter individually and assumed the priors to be mutually independent. For each of the means and their correlation we chose a uniform distribution, while for the variances we selected an inverse Gamma distribution with parameters 0.25 and 1, which yields a very broad distribution over the positive real axis with a peak at 1.

It is possible to fit the hierarchical model using the same procedures we applied to the orginal model. We skip optimization and Frequentist analysis here though. Instead we directly sample the posterior using Gibbs sampling \citep{Casella1992} with parameter groups for each subject and one group for the hyperparameters, sampling each marginal distribution using the Metropolis Hastings algorithm. Specifically, we first cycled through each subject performing one Metropolis Hastings sampling step for the corresponding two individual parameters. Next, we performed one Metropolis Hastings step for the parameters of the Gaussian distribution, which was assumed for the parameter distribution in the population. All proposal distributions were Gaussians with diagonal covariance matrix, adjusted by hand to approximately achieve $25\%$ acceptance rate, and variances roughly proportional to the posterior variances of the parameters \citep{Gelman1996}. We used the same proposal distribution for each subject. Gibbs sampling is especially efficient for hierarchical models, since sampling the parameters of each subject requires only the likelihood for the data of that subject. Thus a whole sweep is computationally only as costly as single likelihood evaluation for updating all parameters. We sampled 3 chains of 10 000 sweeps through the parameters each starting at the maximum a posteriori estimates over all data. As burn in we removed the first 1 000 samples of each chain, which seemed sufficient after visual inspection of the chains. This yielded an effective sample size between 347 and 4472 for the different parameters and the chains seemed to have converged according to visual inspection of the chains and the $\hat{R}$ statistic which had an upper CI bound of 1.06 or less in all cases.

The results of the hierarchical model analysis are shown in Figure \ref{fig:Hierarchical}. First in A, we observe that different subjects are fitted by considerably different sizes for both $\sigma_A$ and $\sigma_F$ and that the estimates for the two parameters are highly correlated, i.e., subjects who have a larger fitted attention span also have a larger fitted inhibition area. Second in panel B we show that the mean saccade length predicted by the model depends strongly on $\sigma_A$ and consequently on $\sigma_F$, as they are highly correlated. Finally we compare the measured mean saccade length to the mean saccade length predicted by the fitted model by simulating as much data as measured for each subject with their posterior mean parameters. The two observables are strongly related, indicating that varying the two spans in the SceneWalk model could account for the difference in saccade length between subjects. Additionally we can observe that the predicted mean saccade length grows with a slope slightly smaller than 1 with the measured saccade length, indicating a slight regression to the mean, as expected and intended for a hierarchical model. 

Looking at the individual subject estimates more closely, we can observe that most subjects (30 of 35) fall into a large cluster, with slightly smaller $\sigma_F$ than $\sigma_A$. However, three subjects have larger fitted inhibition spans and two subjects have extraordinarily large attention and inhibition spans.

\section*{Model comparison in the likelihood approach}
The likelihood concept can be used as a general approach to evaluate how well a given model fits experimental data. Thus, it is possible to compare different models. For likelihood-based comparisons between models one usually assumes fitted parameters. Thus one uses the maximum likelihood, i.e., the best likelihood value a model can reach on the data, when the model's parameters are optimally adjusted. In the following, we denote the maximum likelihood as $L(M) = \max_{\boldsymbol\theta} L_{M}(\boldsymbol\theta|{\rm data})$. 

For the comparisons that we will carry out below, it is important that the log-likelihood is always a relative measure, since it depends on the grid for the observation of fixation positions, the size of the dataset and other dataset specific aspects. Therefore, only the log-likelihood-ratios between models can be compared between different datasets, models, or viewing conditions. Given a null model $M_0$, which defines a reference point, one can compute a likelihood ratio $\Lambda$ to compare a model $M_1$ to the model $M_0$, i.e., 
\begin{equation}
\Lambda(M_1) = \frac{L(M_1)}{L(M_0)}  \;.
\end{equation}
The likelihood ratio $\Lambda$ informs about how many times more likely the data are generated by model $M_1$ than by model $M_0$. For theoretical considerations and for most computations the log-likelihood ratio $\lambda$ is a better choice,
\begin{eqnarray}
\lambda(M_1) &=& \log (\Lambda(M_1)) 
= \log \frac{L(M_1)}{L(M_0)}\\
&=& \log(L(M_1))-\log(L(M_0)) \;.
\end{eqnarray}
The log-likelihood ratio is additive and can be interpreted in a straightforward way, e.g., if $M_2$ is one bit better than $M_1$, which is one bit better than $M_0$, then $M_2$ is two bits better than $M_0$ and the data are 4 times more likely under model $M_2$ than under model $M_0$. 

Also, the log-likelihood ratio can be interpreted in information theoretic terms as the \emph{information gain} about the data generated by the new model compared to the information explained by the original model. Thus the log-likelihood ratio measures how much communication could be saved when specifying a sequence of fixations using a code based on the model. As information theory is well developed \citep[for an introduction]{Ash1990}, it provides a strong theoretical background for log-likelihood ratios in model comparisons. 

In principle likelihood ratios measure the relative quality of the model fits. However, models tend to fit aspects of the data which are purely random, a phenomenon known as \emph{overfitting} \citep[e.g.,][]{Dietterich1995}. Overfitting is the main reason why \emph{model selection}---to which \cite{Zucchini2000} gives an introduction for psychologists---should not be done by directly comparing the likelihoods based on the data used for fitting the models \citep{Myung2000a}. Ultimately the goal of model comparison approaches is to compare the expected likelihood on new data, not on the data used for fitting. Proper model selection and comparison methods are especially critical for comparing models which differ in their flexibility. More flexible models always explain more details of the dataset they are fit to, and thus produce larger likelihood values for the dataset they are fit to. However, more flexible models should only be preferred if the additionally explained details generalize to new data. 

There are two popular quantities model comparison techniques try to estimate and use for comparing models. The first one is the \emph{out-of-sample-prediction error} \citep{Gelman2013}, i.e. one tries to estimate the likelihood of the parameters fitted on the given data on a new dataset. The second one is the \emph{evidence} for a model which is the denominator of the Bayesian formula---$\int_\Omega P(\boldsymbol\theta)L(\boldsymbol\theta|{\rm data})d\boldsymbol\theta$---i.e. the total probability to observe the data according to the model with the given prior $P(\boldsymbol{\theta})$. For a new dataset this means the evidence estimates the models performance using only the prior information about the parameter value. Consequently the evidence critically depends on the prior and can be arbitrarily bad if the prior assigns large probability to parameters with low likelihood. The ratio of evidences for two models is called the Bayes factor.

The first approach for model selection are metrics which add a correction or penalty term for more flexible models. These metrics are generally called information criteria and are usually formulated in terms of the \emph{deviance} ($-2\lambda(M)$)---a general measure of prediction error---which is directly computed from the likelihood and contains exactly the same information, but reverses the sign. Thus smaller information criteria correspond to better models.

Classical examples for this procedure are the Akaike Information Criterion \cite[AIC,][]{Akaike1974} and the Bayesian Information Criterion \cite[BIC,][]{Schwarz1978}. The AIC was formally introduced as a first model selection criterion, defined as: $AIC(M) = -2\lambda(M) +2 \operatorname{dim}(M)$\footnote{\label{symbolMeaning}$\operatorname{dim}(M)$ representing the dimensionality of the model, i.e. the number of parameters, $n$ the number of independent observations}. It represents a simple large sample bias correction obtained from Fischer information theory estimating out-of-sample-prediction error. The BIC \citep{Schwarz1978} was introduced as an approximation to the evidence in favour of a model in the case of an exponential family model. Thus it effectively aims to estimate the generalization quality to new data which requires new fitted parameters. For $n$ independent observations it is defined as\footnote{The original criterion was half the value described here. However the version reported here seems to be the more commonly used one today.}: $BIC(M) = -2\lambda(M) +\log(n) \operatorname{dim}(M)$\footnotemark[4]. This obviously does not contain the prior and is a coarse approximation to the evidence. From very small datasets on this penalty will be larger for the BIC than for the AIC, e.g. the BIC will prefer parsimonious models more strongly than the AIC corresponding to the harder generalization task estimated by BIC. 

The classical information criteria---AIC and BIC---both result in very small corrections of the raw likelihood. Our dataset contained 13908 and 13306 fixations for natural images and texture images respectively. Thus for our model with 8 free parameters the AIC and BIC penalties would maximally be $0.0008\frac{\rm bit}{\rm fix}$ and $0.0041\frac{\rm bit}{\rm fix}$ respectively, while the differences between models are much larger. In contrast, our cross validation results below suggest that the actual difference between fitted data and new data is much larger. Thus AIC and BIC seem to provide bad estimators in our case of complex dynamical models. 

Very similar Bayesian evaluations exist \citep{Gelfand1994}, which estimate generalization of the posterior predictive distribution instead of generalizations based on a point estimate for the parameters. Nonetheless, the aim stays to predict how likely new data will be according to the model.

Fortunately direct formulas to approximate model performance in fully Bayesian terms from sampling results exist \citep{Gelman2013}. Thus a Bayesian Model comparison is possible, once a representative sampling is available for the posterior on the parameters of each model. Examples for this approach aimed at generalization to new data from the same parameters are the Deviance Information Criterion \citep[DIC,][]{Spiegelhalter2002} which approximates the posterior as the mean estimate and the Widely Applicable Information Criterion \citep[WAIC,][]{Watanabe2010}, which directly uses the sampling estimate for the posterior predictive. Both these criteria also use the posterior samples to their advantage to produce a more accurate estimate for the out of sample prediction quality. Similarly, there is also a Bayesian alternative to the BIC, the Widely Applicable Bayesian Information Criterion \citep[WBIC,][]{Watanabe2013}.

Calculation of the Bayesian information criteria requires an estimate for the posterior distribution on the model parameters, i.e., a sampling of the posterior. As we compare 10 models below and only have a sampling for one of these models, we do not perform these analyses here. However, such analyses should be considered especially when one studies other models like hierarchical models for example for which cross validation is not straight forward. And of course, once the posterior predictive is used for prediction, this should be the measure to be compared in the cross validation.

One should note that the penalties of all information criteria per data point (i.e., fixation or scanpath) converge to zero for growing dataset size. Thus larger datasets will raise a preference for more detailed models if there is any advantage for prediction. This makes sense as the criteria penalize complexity only when this complexities cannot be calibrated well enough to improve predictions with the given data \citep{Burnham2004}.

\begin{figure*}
\includegraphics[width=\textwidth]{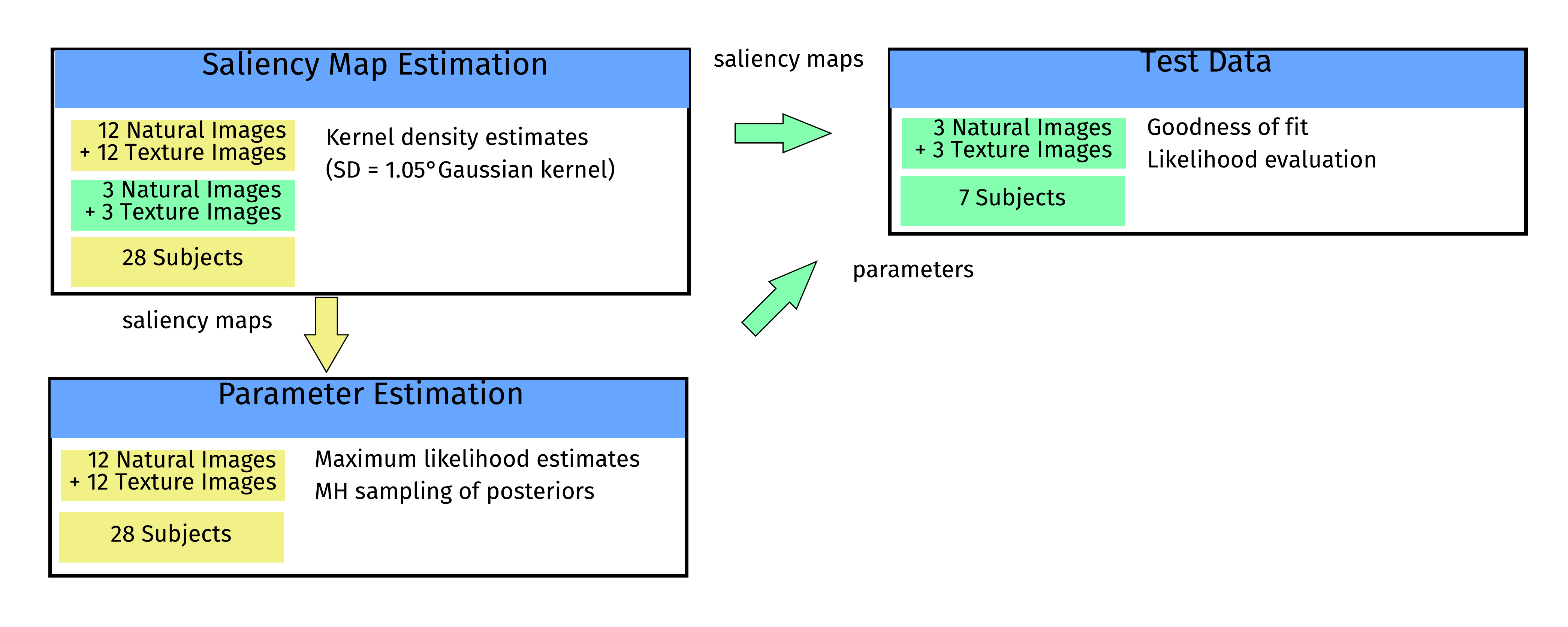}
\caption{To guarantee that the model is fit to a different dataset than the one used for evaluation many possible separations exist. Here we display the separation of our dataset into training and test data used for each fold of cross validation. Data from 28 human observers on $2\times 12$ images (yellow) were used for parameter fitting, while the data from 7 different observers on $2\times 3$ test images were used for model tests (green). \label{fig:CrossValidation}}
\end{figure*}

A different more data driven approach to estimate the quality of out of sample predictions is \emph{cross validation}, which is frequently used in machine learning, but has been introduced to the psychological literature as well \citep{Browne2000}. For cross validation the dataset is split into $n$ subsets. Then the model is fitted to $n-1$ of the subsets---the \emph{training set}---and evaluated on the one subset not used for fitting---the \emph{test set}. This is repeated for each of the subsets being the test set and the results are averaged. This procedure applies to Bayesian and Frequentist evaluation equally, but is more frequently used with point estimates and Frequentist evaluation. 

For dynamical models for eye movements in scene viewing, two separate factors induce variability for which overfitting could occur: human observers (subjects) and stimuli. To avoid problems of overfitting for these two factors, we split our data across both factors and perform 5-fold cross validation using splits into training and test set as illustrated in Figure \ref{fig:CrossValidation}: For each fold we used the data obtained from 28 subjects on 12 natural images and 12 texture images for \emph{training}. For evaluation we run the model on data obtained from 7 other subjects on 3 other natural images and 3 other texture images. To compute the empirical saliency maps, we used the 28 training subjects on both training and test images. There are also  data for the training subjects on the test images and the test subjects on the training images, both of which are not used here to completely isolate training and test sets from each other.

For each fold we fitted the model to the training data using the genetic algorithm of MATLAB with settings as for the original fitting process on all data described above. However we noted that there was exactly one more local maximum to be found at small ($\sigma_F\approx.5^\circ$), fast ($\omega_F\geq 10$) inhibitions, to which the genetic algorithm converged for some folds. To find the global maximum in every case nonetheless, we started a subsequent fminsearch optimization from each of these 2 maxima for each fold and took the better one as the global maximum. In all folds and all models the global maximum had similar sized attention window and inhibition and generally similar parameter values to the fit of the subtractive model to all data described above. The other local maximum was usually around 1000 worse on the log-likelihood scale for the training data. Thus the decision was always clear cut. Nonetheless this additional local maximum can be understood. Effectively it implements an inhibition for saccade targets very near to the current fixation. Saccades to these targets would not be detected as such by the data preprocessing such that such short saccades indeed do not occur in the dataset and cannot occur in a dataset. Thus this model adaptation indeed would be predictive, but not informative about any underlying processes of eye movement behavior. 

\subsection*{Results on model comparison}

\begin{figure}
\unitlength1mm
\includegraphics[scale=.65]{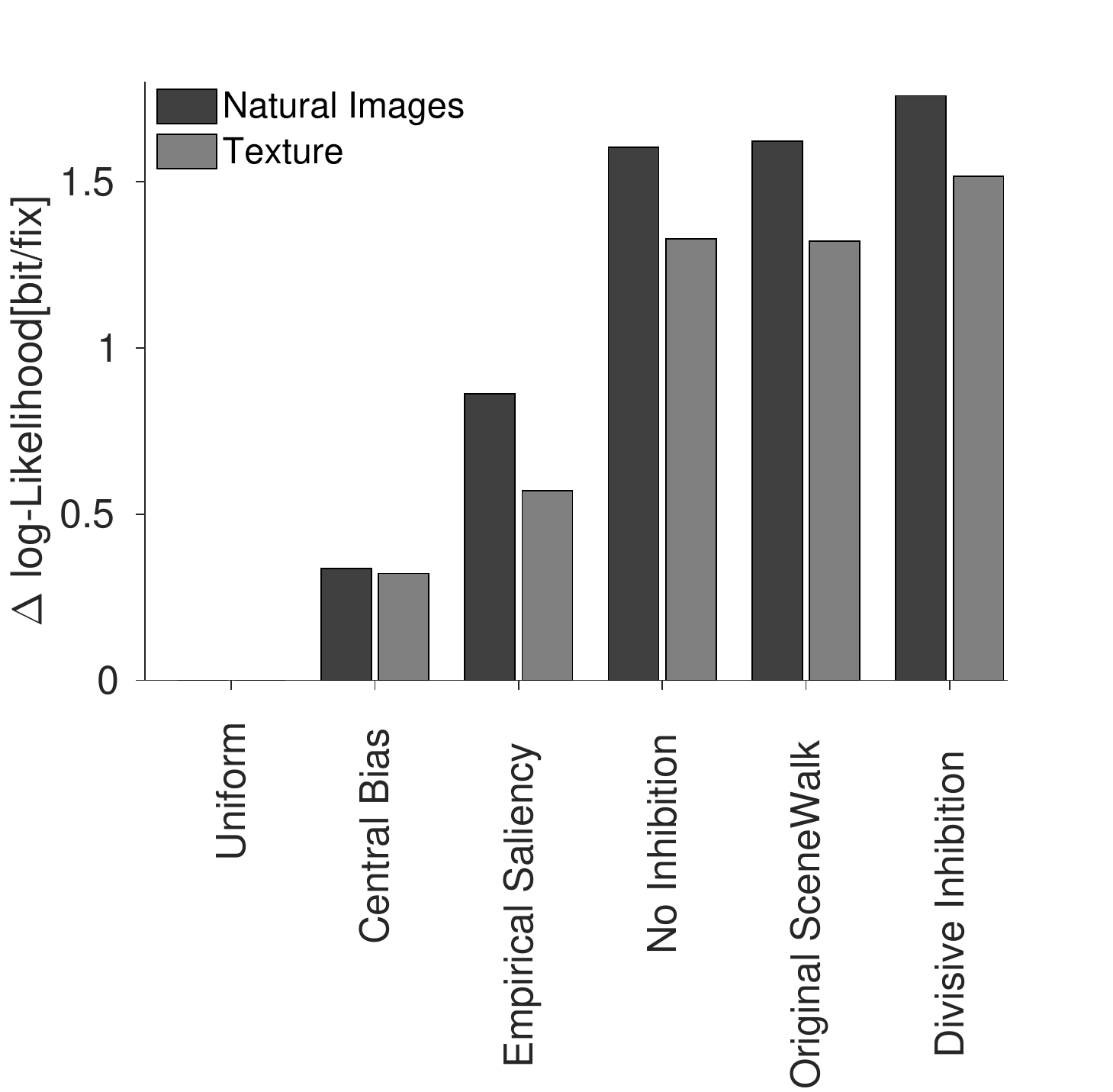}
\caption{Bar plots for the models' log-likelihood differences to the uniform distribution null model. We split here by the two experimental conditions, which differed in the images presented. For the texture models the density map is much less informative than for the natural images. The {\sl central bias/central fixation bias} model is a kernel density estimate from the fixations on all other images. The {\sl empirical saliency} is the kernel density estimate from the fixations of other observers on the same image. Finally, {\sl No inhibition, Original SceneWalk} and {\sl Divisive Inhibition} refer to the three variants of the {\sl SceneWalk} model, which we investigate in detail here. \label{fig:ModelCompare} }
\end{figure}

To perform our comparison we split the data as explained above, fitted the model to each of the 5 training sets and computed the log-likelihood of each model on each test dataset. Then we divided the resulting likelihood value by the number of fixations to normalize the results regarding the size of the dataset. Thus we measure all differences in bits per fixation $[\operatorname{bit}/\operatorname{fix}]$. According to this null model, the uniform distribution over the whole image distributes a probability of $2^{-14}$ for every fixation to each grid point, since we calculated all maps on a $128\times 128$ grid. This results in a log-likelihood of $-14\,{\rm bit}/{\rm fix}$. We ran separate evaluations for texture images and object-based natural scenes presented in the experiments; the log-likelihoods are plotted in Figure \ref{fig:ModelCompare}. Overall, we find a gain for the empirical saliency model over center-bias prediction and a considerable gain in likelihood for the SceneWalk model. 

The information gain for the saliency model differs strongly between natural textures and natural scenes, which was expected as the gaze patterns over texture images were more uniform than the corresponding data for natural scenes. This difference carries over to our dynamical model, as this uses the empirical saliency as an input predicting where human observers want to look. However, the increase in likelihood due to the dynamical principles is comparably large for texture images and for scenes. This result lends support to the view that the same dynamical principles of scanpath generation are underlying texture images and natural scenes. 

We also evaluated the model with the parameters values fitted by \citet{Engbert2015}. This yields a likelihood value of $-12.96$~${\rm bit}/{\rm fix}$ for natural images and  $-13.10$~${\rm bit}/{\rm fix}$ for texture images for the training data (not shown in the figure). This indicates that the model explained the data better than empirical saliency even with the parameters not optimized for the likelihood. However, with the new parameter values the model generates higher likelihood values per fixation on the test sets it was not trained on (natural scenes: $-12.38$~${\rm bit}/{\rm fix}$, textures: $-12.68$~${\rm bit}/{\rm fix}$).

To compare different model specifications against each other, we generated two new model variants---one without inhibition and one with divisive inhibition---described in detail above. Additionally we questioned whether the introduction of the exponents $\lambda$ and $\gamma$ were necessary. To test this we generated model variants with one or both of the exponents fixed yielding 4 variants of the subtractive original SceneWalk model, 4 for the divisive model and 2 for the model without inhibition.

\begin{figure*}
\unitlength1mm
\includegraphics[scale=.7]{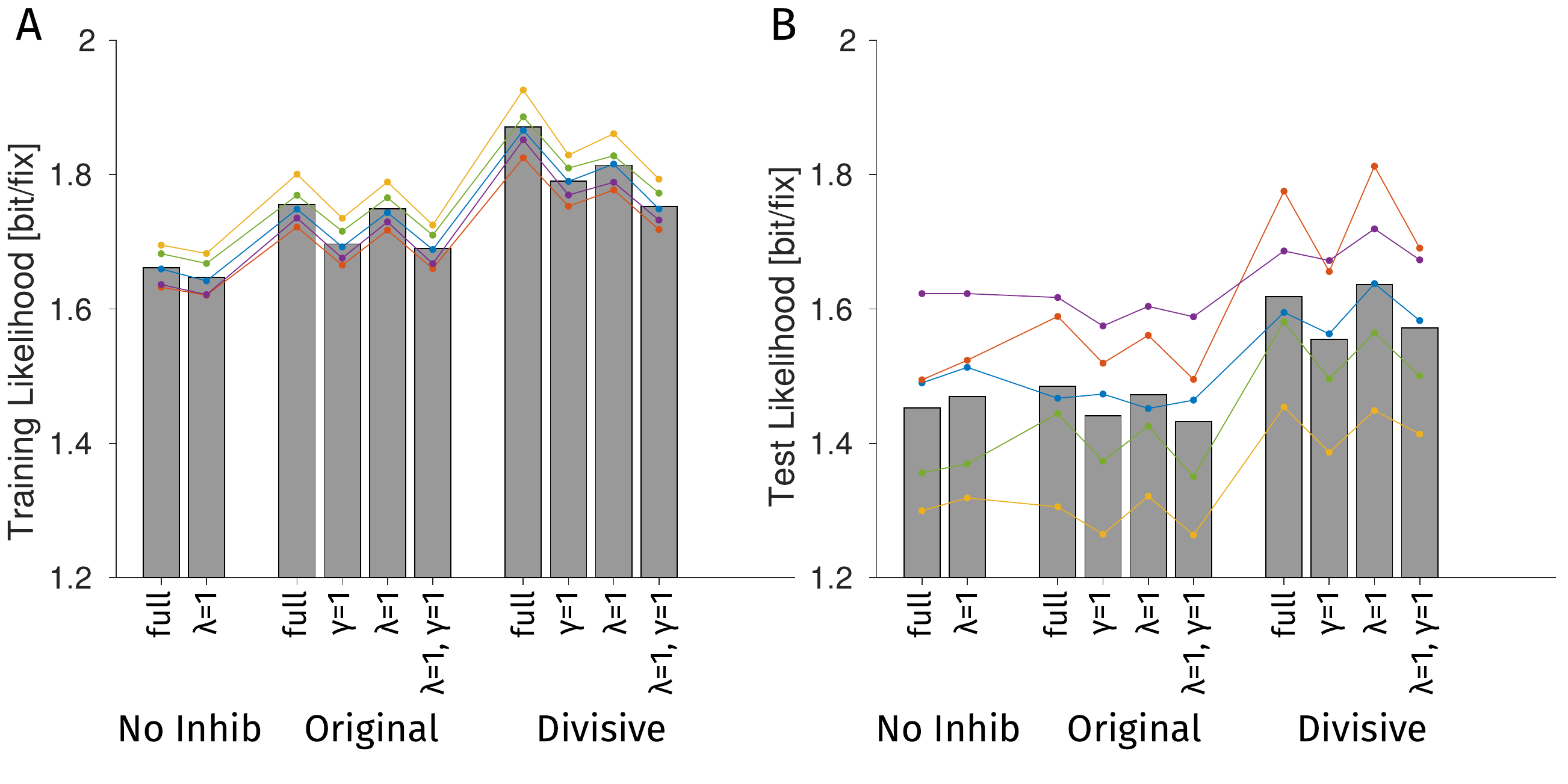}
\caption{Bar plot comparing log-likelihood differences to the uniform distribution null model, exploring the effects of the exponents. Each bar is the average test set performance of the 5 folds of our cross validation procedure. The colored lines plot the results for the 5 folds. \textbf{A:} The likelihoods on the training datasets, which should not be used to judge the models, but are informative, whether the model fitting worked properly. \textbf{B:} The likelihoods on the test datasets, which can be used to compare models. \label{fig:ModelCompareDetail} }
\end{figure*}

First, as a check on the results it is informative to look at the performance of the models on the training data, we display in Figure \ref{fig:ModelCompareDetail}A, although these values should not be used for model comparison. Evaluated on the training data a model which contains another model as a special case must be at least as good as the contained model on each of the training sets. This sanity check was how we first noticed that some of the optimizations had ended in a different, wrong local maximum. Also comparing the training set and test set results provides some insight how substantial the flexibility problem is for the specific model.

The test set results of these more detailed comparisons are displayed in Figure \ref{fig:ModelCompareDetail}B. We find that overall the divisive inhibition model provides the best performance followed by the original SceneWalk model and finally the model without inhibition. Within each model type the exponent $\gamma$ seems to improve the model fit, while the fits with free $\lambda$ yield equally good performance or even worse performance than fixing $\lambda=1$ (using the attention map without non-linear distortion). The model to choose from our pool is thus the divisive inhibition model with a large, fitted $\gamma$ and $\lambda$ fixed to 1.

Note that all the models with inhibition have a qualitatively similar behavior and typically computed statistics on scanpaths cannot discriminate these models, as we discuss below. Thus the likelihood based comparisons allow us to differentiate models we could not differentiate otherwise. A restriction of these model comparisons is, however, that they do not come with a measure of uncertainty like standard errors, credible or confidence intervals or adequate statistical tests\footnote{Some classical $\chi^2$ tests of model fit exist. As they are based on the same approximations as the AIC and BIC, we doubt that they produce correct conclusions here.}. Thus we cannot provide a hard statistical measure how sure we are about the order of the models although the differences can be interpreted in size.

\section*{Goodness-of-fit for specific measures and spatial statistics}
While we used the likelihood as a general measure of model fit to experimental data, the likelihood remains a relative (i.e., depending on a null model) and global measure (i.e., no specific statistical properties are addressed). Thus, there are at least two reasons to check other statistics additional to performing a likelihood-based approach to parameter estimation or model comparison. First, to analyze the absolute performance of the model, and, second, to understand which aspects of the data are modeled adequately and which other aspects are modeled poorly. 

The first reason, judging the absolute quality of models, is to check that they are good enough to be interesting, which is subsumed under \emph{goodness-of-fit} analysis in statistics \citep{Pitt2002,Wichmann2001}. In statistics, the importance of goodness-of-fit analyses is emphasized, since the theory of parameter estimation for models is built on the assumption that there is a correct solution, i.e., model parameter values exist that actually  generated the data. So, if a model cannot explain the data well for any parameter value, the best estimate for the parameter might be meaningless, even when the best parameter value is defined by generating the highest likelihood for a given model. For the same reason, Bayesian inference methods may fail if there are no good models in the set assumed a priori.

\begin{figure*}
\unitlength1mm
\includegraphics[width=165mm]{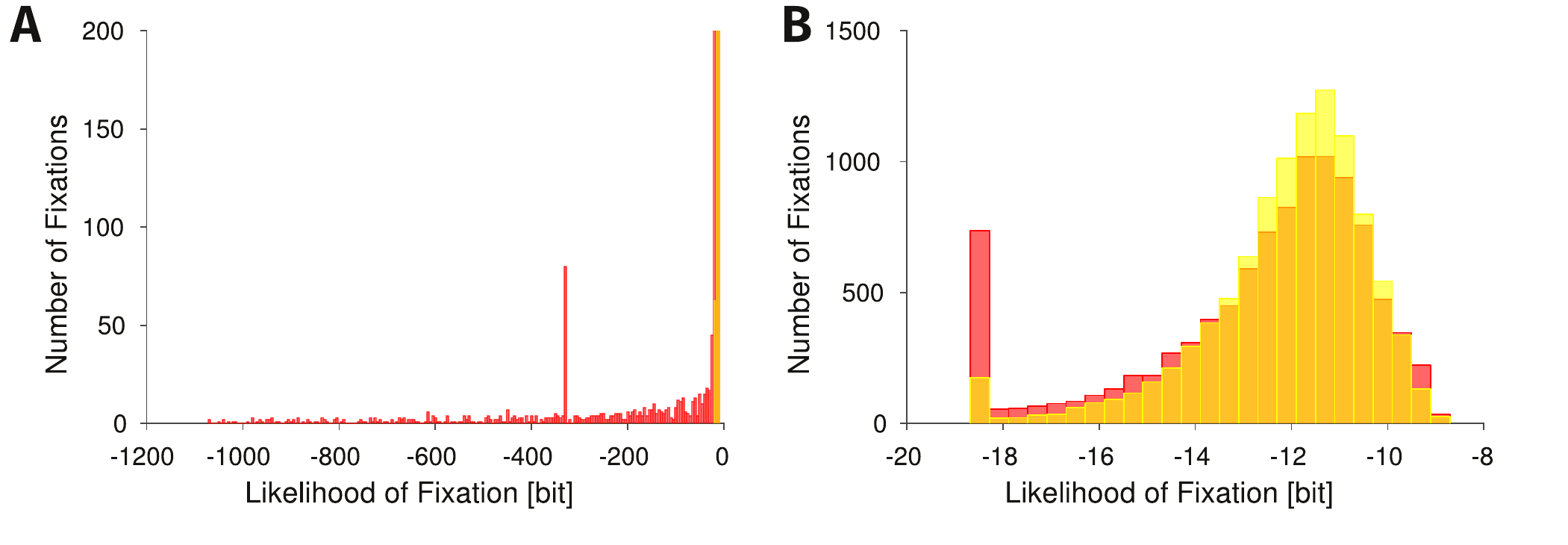}
\caption{Histograms of the likelihood of individual fixations on the test dataset (red) and on data generated from the model (yellow) {\bf A}: Employing a model without mixing with a uniform distribution (setting $\zeta=0$ in Eq.~(\ref{eq:pidef})). The considerable number of extremely unlikely fixations led us to include the mixture with a uniform distribution in Eq.~(\ref{eq:pidef}). {\bf B}: Employing the full model with the mixture, extremely unlikely fixation positions no longer occur. \label{fig:LikelihoodHist}}
\end{figure*}

To get an idea about the absolute quality of the model's predictions for data, the easiest way is to simulate data by the model and to compute statistics for these data in exactly the same way as it is done for the interpretation and statistical analysis of experimental data. A comparison of the resulting statistics gives a good indication of the quality of the model's fitness. 

Based on the likelihood it is also possible to test how (un-)likely the measured data are, compared to the expected likelihood of data from the model. This expected likelihood can be computed by simulating larger amounts of data from the model and computing its likelihood. For a perfect fit, the measured data should have a similar likelihood as datasets simulated from the model, which represents a test whether the model's output variability matches the variability of the observed data.

We performed such an analysis by simulating as much data as we had collected and computed the likelihood of this data. We compare histograms over the log-likelihood per fixation for simulated and experimental data in Figure \ref{fig:LikelihoodHist}. First, in Figure~\ref{fig:LikelihoodHist}A, we ran the analysis on a model without the mixture with a uniform distribution, i.e., choosing $\zeta =0$. According to this model some of the observed fixations were extremely unlikely, i.e. the model predictions were to specific, which motivated us to include the mixture with a uniform distribution. In Figure~\ref{fig:LikelihoodHist}B, we show a histogram of the log-likelihoods for the full model, again for the measured data and simulated data from the model. For the full model, the mean log-likelihood of the simulated data is $-12.11$~${\rm bit}/{\rm fix}$, $\Delta = 1.89$~${\rm bit}/{\rm fix}$ (raw value, difference $\Delta$ to a uniform distribution), which is roughly equal to the likelihood for the training data of $-12.08$~${\rm bit}/{\rm fix}$,~$\Delta=1.92$~${\rm bit}/{\rm fix}$, but larger than for the test data for which the model reaches only $-12.67$~${\rm bit}/{\rm fix}$,~$\Delta=1.33$~${\rm bit}/{\rm fix}$. The small difference between training data and model-generated data suggests that the model did not overfit the data dramatically, i.e., we would expect the model to be roughly as good as it is for the data, if the data were generated by the model. The difference between training and test data suggests that the model does not generalize to the test dataset perfectly, which is mainly caused by an increased number of highly unlikely fixations (Fig.~\ref{fig:LikelihoodHist}B). It seems plausible that these are fixations in regions where none of the observers in the training set fixated (regions of low empirical saliency). This indicates that a higher number of observers for estimating the empirical saliency map would be beneficial to our approach.

The second motivation for additional model analyses is to decide which aspects of the data are modelled well, and which are not described adequately. It is important to further improve models and to choose appropriate models for different situations and modelling goals. Generally, measures used for this analysis should be interpretable for the modeller and other researchers. Some more detailed information can also be extracted from the likelihood calculations as this calculation is split over the different observations. Thus for each individual observation a separate likelihood can be computed and one can check which measured scanpaths or individual fixations are especially likely or unlikely according to the model providing some additional, more specific information. 

For the SceneWalk model we started with an analysis of standard statistics from eye-movement experiments. As a first step, we compared the overall fixation density of model and data. To quantify the comparison, we computed the \emph{Kullback Leibler Divergence} (KL-divergence) of the fixations predicted by the model against the fixations made in our experiment. This standard measure is computed as 
\begin{equation}
KL = \int_I p(x) \log\frac{p(x)}{q(x)} {\rm d}x \;,
\end{equation}
where the integral is computed over the full image $I$. 

The fixation density generated by the model does not fit the empirical saliency perfectly, but perturbs it slightly through its dynamics. However, the predicted distributions diverge less from the true density (average KL-divergence = $0.1997$) than any saliency models, which minimally reach 0.54 and 0.37 for the two datasets in the MIT saliency benchmark \citep{mit-saliency-benchmark}. The good performance of the SceneWalk model is not surprising here, since we used the empirical fixation density as an input to our model.

\begin{figure}
\unitlength1mm
\includegraphics[width=85mm]{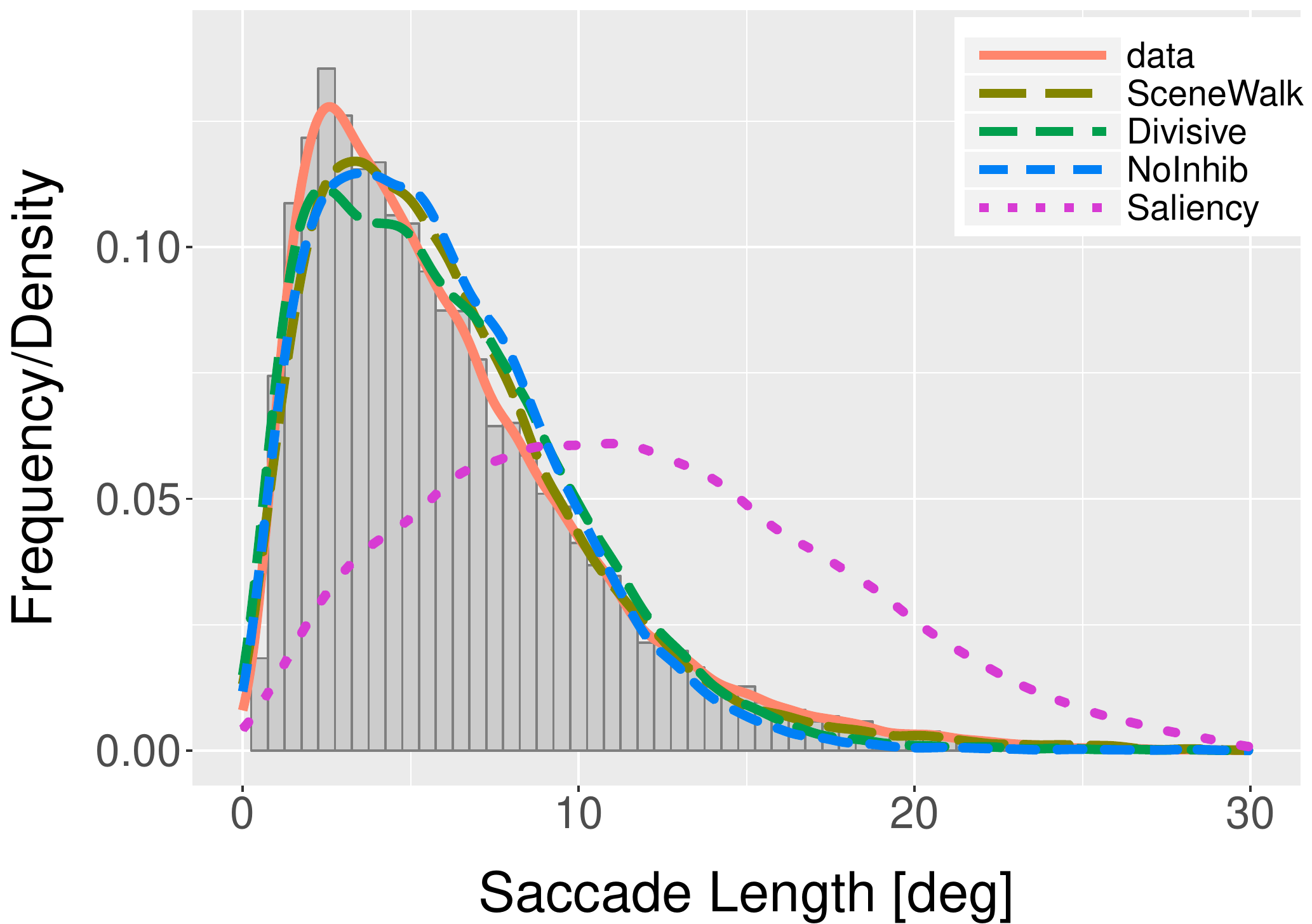}
\caption{Comparison of model and data based on saccade lengths. The plots present the saccade length distribution over all images for experimental data and model simulations. \label{fig:SaccLength}}
\end{figure}

Next, we looked at the distribution of the saccade lengths, a first aspect of the model dynamics. The results of this analysis are given in Figure \ref{fig:SaccLength}. The saccade lengths in the model and data are very similar and the variance over images is small in both model and data,  while the variance over subjects is substantial as we discussed above. Also the competitor models without inhibition and with divisive inhibition fit the distribution of saccade lengths well such that the saccade length distribution does not clearly differentiate these models from each other. However, simply drawing fixations independently from the empirical saliency map yields an entirely different, wrong distribution.

\begin{figure}
\unitlength1mm
\includegraphics[width=85mm]{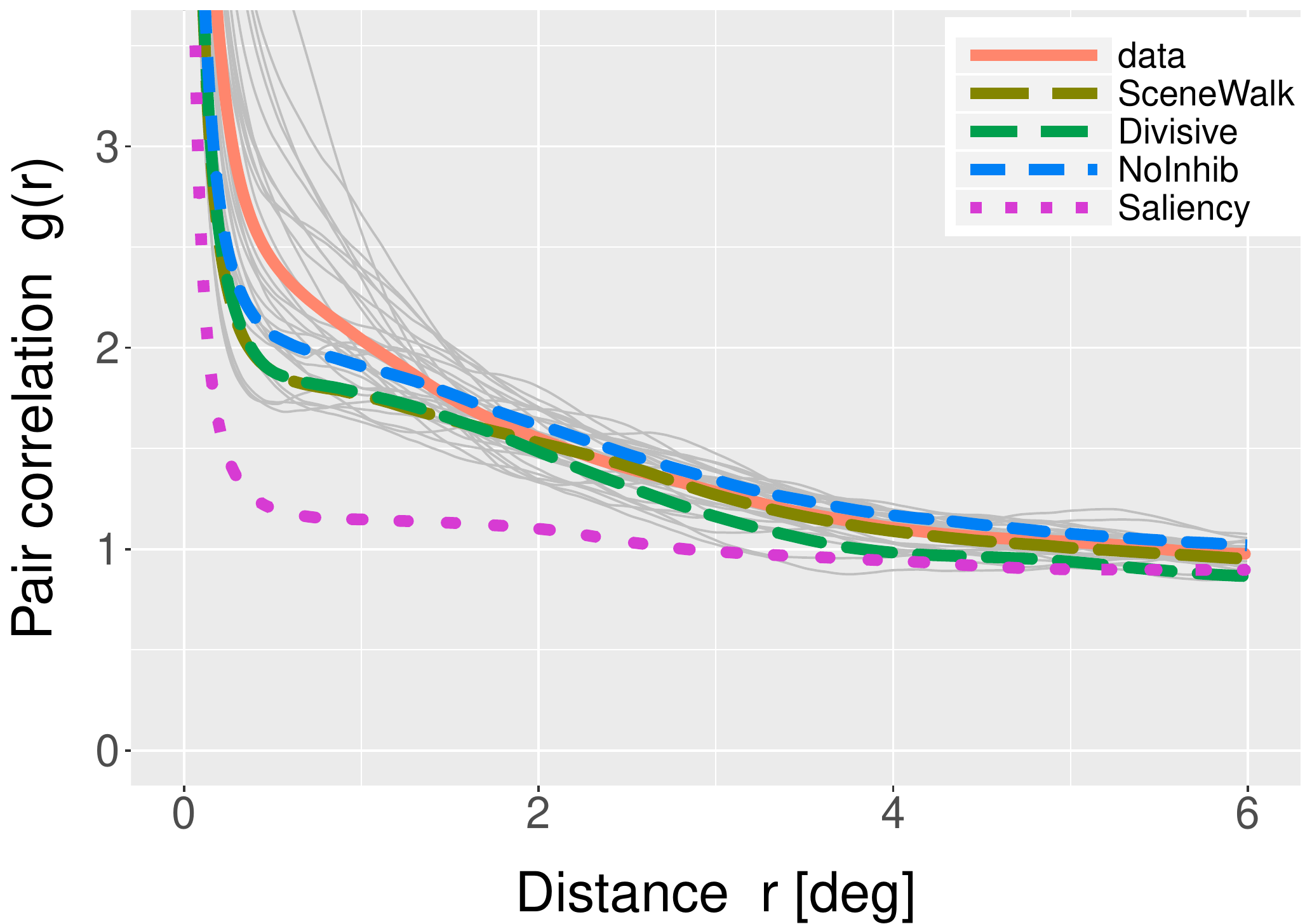}
\caption{Comparison of models and data based on the pair correlation function (PCF). The mean PCF for each of the models is plotted in color. For the data the mean is shown in color as well and the pair correlation functions for individual images are plotted in gray. Higher values than one indicate clustering or aggregation, i.e., fixations at distance $r$ are more abundant than expected on average from independently drawn fixations from the fixation density. Values smaller than one indicate repulsion, i.e, fixations at distance $r$ are rarer than expected for independently drawn fixations. \label{fig:PCF}}
\end{figure}

Recently, methods from the theory of spatial point processes were introduced into the analysis of fixation patterns in scene viewing \citep{Barthelme2013,Engbert2015}. Most of the standard statistical measures are {\sl first-order statistics}, e.g., the 2D density of fixations. 
For the SceneWalk model, we computed the {\sl pair correlation function} \citep{Engbert2015} as an example for a second-order spatial statistic. The pair correlation function (pcf) describes how frequently two fixations with a certain distance occur in one scanpath normalized against the frequency expected for a random selection from the fixation density. Values higher than one indicate that fixation patterns are more aggregated than could be expected from the first-order spatial inhomogeneity of the process. As the pair correlation function includes later returns to earlier fixated positions, this function measures a different property than the saccade length distribution. In experimental data, the pair correlation function usually indicates a clustering at small distances below $3-4^\circ$ \citep{Engbert2015}. Comparing the pair correlation functions estimated from experimental data and model predictions in Figure \ref{fig:PCF}, it is obvious that all models fit the pair correlation function much better than a simple random process that draws fixations from the empirical density map. However this measure seems not to differentiate between the different types of inhibition either.

\section*{Discussion}
The key motivation for the current study was to apply the likelihood approach to the evaluation of dynamical cognitive models and, in particular, for model parameter estimation and model comparison. Dynamical cognitive models are formulated by evolution equations (temporally discrete or continuous) and evaluated against time-ordered data (time series). As a specific example, we investigated the problem of saccade generation, where the dynamical model determines the probability $\pi(x,t)$ to select a saccade target position $x$ at time $t$. In the SceneWalk model \citep{Engbert2015}, this probability is computed from activation fields at any point in time. Thus, we can compute the corresponding probability for a fixation and force the model to generate the gaze shift to the new fixation position. This procedure of direct computation of the likelihood will work for the broad class of dynamical models that generate continuous-time activations for the prediction of discrete behavioral events \citep{Erlhagen2002}.

For the interpretation, we normalized the likelihood with respect to the number of fixations in a given dataset to obtain a measure that is independent of the size (length) of the fixation sequence. Furthermore, we suggested to compare the likelihood to the likelihood obtained from a uniform distribution to get a measure which is independent of grid and image sizes. For simpler, non-dynamical models this comparison to chance performance is a standard procedure. Additional non-dynamical models were used to generate likelihoods to compare to the dynamical model. Such non-dynamical density models \cite[e.g., the central fixation bias,][]{Tatler2007} represented a convenient statistical baseline for our computations. Finally, we investigated two variants of the SceneWalk model to show that the likelihood can be applied as a powerful tool to distinguish  different dynamical models with highly specific assumptions.

The likelihood as a global measure of model performance can be used as a tool for the estimation of model parameters. Fitting models based on the maximum likelihood concept has a long tradition in statistics and some clear advantages over other parameter fitting procedures, including mathematical proofs for the convergence and sufficiency of the parameter estimate. A practical advantage is that the likelihood is a scalar value, which does not rely on simulating complex discriminating statistics. 
Additionally, model fitting based on the likelihood is the starting point for Bayesian inference about parameter values, which provides new insights to other parameters that could explain the data and, thus, statistical comparisons on whether the parameters differ between datasets or conditions. 

For the SceneWalk model \citep{Engbert2015} we computed parameter values using maximum likelihood estimation and sampled the posterior for Bayesian parameter estimation. This parameter estimation technique allowed us to fit all the parameters of the model, which was impossible in the original publication. The parameters found by optimizing the likelihood reproduce all the statistics the original publication reported, while the parameters from the original publication perform significantly worse in terms of likelihoods. Additionally, we computed a full posterior probability over the parameters that informs about which parameters are constrained by the data well and which parameters are not constrained by the data.

Furthermore, the likelihood-based evaluation helped us to improve the original model. Using a hierarchical model, we found that the known differences between subjects in their average saccade length \citep{Castelhano2008} could be fit well, by allowing the size of the attention window and the size of the inhibition to vary between subjects. Furthermore likelihood based comparisons between models allowed us to show that the dynamics and the inhibition both improve model predictions. And additionally we could differentiate different variants how the excitatory and inhibitory maps are combined. For experimentally-motivated statistics, these specific model variants made very similar predictions. Among the models analyzed here, a divisive inhibition model with a fixed numerator exponent $\lambda$ seems to fit the data best---and even better than the original SceneWalk model.

With the SceneWalk model, we focus on fixation locations and take fixation durations as given (or a random process with given mean and variance). This is, however,  not necessarily a restriction of the likelihood approach. Models which compute probabilities for fixation durations \citep[for example]{Nuthmann2010,Trukenbrod2014} or for both the durations and locations of fixations could be fit and evaluated using the same techniques we present here for locations only. There are recent studies on fixation durations for scene viewing \citep[e.g., ][]{Laubrock2013}. Furthermore, the prediction of fixation durations is a main aim for models of eye movements during reading \citep{Reichle2003, Engbert2005}.

In this article we used relatively simple gradient free optimization algorithms and the Metropolis-Hastings algorithm for their conceptual simplicity, which eased the presentation. However, there might be more efficient algorithms for solving the optimization and sampling problems in the SceneWalk model and certainly different algorithms will be best or easiest to implement for different models. Also, the optimizations and samplings for complex models may take hours, days or even months of computation time. Thus efficiency is important as it may make the difference whether an analysis is feasible with given computational resources or not. Consequently, it can be worthwhile to invest some time to try different optimization algorithms including global optimization algorithms, when local minima are a problem. Similarly there is broad literature on how to (adaptively) tune MCMC-algorithms \citep[e.g., ][]{Roberts2009,Andrieu2008,Gelman1996,Haario2001,Haario2006} and efficient sampling algorithms \citep{Robert2009,Robert2013,Brooks2011}.

An especially large step in efficiency for both optimization and sampling can be made if a gradient of the likelihood can be calculated with reasonable efficiency. For optimization highly efficient gradient based algorithms, i.e. quasi-Newton methods like the BFGS algorithm are available. The original gradient based sampling algorithm is the Hamiltonian Monte Carlo (HMC) method introduced by \cite{duane1987} \citep[see][for an introduction]{neal2011}. By now there are many variants of HMC available, including adaptive methods like the No-U-turn Sampler \citep[NUTS,][]{Hoffman2014}, which works behind STAN \citep{Carpenter2016}, one of the most recent general purpose samplers. These samplers contain automatic differentiation tools, which remove the necessity to code a gradient computation by hand. Also independent tools to compute derivatives automatically are able to differentiate virtually any computable function \citep{Tensorflow2015,Theano2016}, which allows computation of a derivative for many models.

As a next step the likelihood evaluation permits comparisons between different models. To avoid overfitting such comparisons were carried out using cross validation. Here, the SceneWalk model \citep{Engbert2015} was compared to a statistical model of the central fixation bias and to a model that sampled fixation positions from the empirical saliency map. We found that the SceneWalk model outperforms the empirical saliency model by $0.75 \frac{\rm bit}{\rm fix}$, which highlights the importance of incorporating influences of previous fixations into predictions for upcoming saccade targets. Consequently, a saliency model alone is not a good models for scanpaths, no matter how closely it matches the fixation density. 

As the likelihood is a relative measure, it is necessary to check whether the fitted model is reasonably good in terms of absolute measures. For the SceneWalk model we demonstrated the adequacy by comparing different summary statistics computed on model predictions to the corresponding statistics obtained from experimental data. We found that the model reproduced the fixation density, saccade length distribution and the pair correlation function with parameters computed via maximum likelihood estimation.

For scanpath models in eye-movement research, the likelihood approach to parameter estimation and model comparison is most interesting as there is no general consensus on a metric for comparing models so far \citep{Pitt2002,LeMeur2013}. Instead, many statistics on specific aspects of scanpaths were proposed, which allow judgements whether a given model shows some specific effects or not. However, a global account of how adequately the model fits the experimental data is currently lacking. We demonstrated that such global measures could be provided by the likelihood approach. 

In the likelihood approach, any scanpath observed in humans must have a probability larger than zero under the model, as the likelihood vanishes otherwise, indicating only that the model cannot explain the data. A second constraint on the model is that the likelihood can be computed. As we showed above, it is sufficient to be able to numerically generate the probability for the next fixation given the previous ones. This is not a strong constraint as most eye movement models on natural scenes even explicitly represent a probability map for the next fixation \citep[for example]{LeMeur2015,Zelinsky2008,Zelinsky2013}. 

We believe that model evaluations based on the likelihood are promising for many other psychological models. Indeed, for some models the evaluation is already routinely done using likelihoods, for example for receiver operating curves \citep{Ogilvie1968}, diffusion models \citep{Ratcliff2002} or psychometric functions \citep{Wichmann2001} and recently for saliency models and fixations on static images \citep{Barthelme2013,Kummerer2015}. 

One favourable aspect of the SceneWalk model is that it is deterministic---there is only a single way for the model to produce time-dependent activation maps for a given sequence of fixations. If there were multiple possible internal states compatible with the observed data, then the computation of the likelihood would require an integration over all possible internal states. Such integration could render evaluations of the likelihood function less effective or even impossible for other models.  For such complex models with many possible internal states and large datasets efficient computational techniques for combined state and parameter estimation have been developed in particular in the field of {\sl data assimilation} \citep{Stuart2015,Reich2015}. Furthermore, processing time-ordered datasets leads naturally to the consideration of {\sl sequential Monte Carlo methods} \citep{Doucet2001,Chopin2013}, to bring computational demands into a manageable range.

For some model classes computation of the likelihood might be too time consuming or the likelihood function too complex for further handling. However, even for such models, mathematically well founded approximations to the likelihood methods were proposed: \emph{Pseudo-likelihood} methods compute an approximation to the likelihood \citep[for example]{Wood2010}. Alternatively, \emph{pseudo-marginal Monte Carlo methods} \citep{Beaumont2003,Andrieu2009} can be utilized which, while involving approximations, can be shown to provide consistent estimates. Here one could also consider replacing the likelihood by an appropriate {\sl scoring function} \citep{Gneiting2007} which provides an alternative metric to rank models in an objective manner. Moreover, {\sl Approximate Bayesian Computation} (ABC) allows an approximation to full Bayesian inference without a likelihood \citep{Turner2012,Wilkinson2013,Barthelme2011,Barthelme2014}. These methods  preserve some of the benefits of the likelihood approach to parameter estimation and model analysis and can even be used to do model selection. For dynamical models this is discussed for example by \cite{Toni2009}.

\section*{Conclusion}
We proposed and studied a likelihood approach for the evaluation of a dynamical cognitive model for the control of saccadic eye movements. The likelihood can be used for parameter estimation and model comparisons as it makes the full range of statistics available, from maximum likelihood estimation through Bayesian estimation and hierarchical models to proper model comparisons. Compared to non-dynamical models, the dynamical model generated a significant increase in predictive power by introducing sequential dependencies. Our approach is a promising tool for the evaluation of dynamical models that predict sequences of discrete behavior (e.g., fixation position, movement onsets) in general and for human scanpaths in particular.  

\section*{Acknowledgments}
We thank Simon Barthelm\'e, Grenoble, for valuable discussions. This work was supported by grants from Deutsche Forschungsgemeinschaft to R.E.~(grant EN 471/13-1) and to F.A.W.~(grant WI 2103/4-1)

\bibliographystyle{apacite}
\bibliography{Likelihood.bib}

\end{document}